\documentclass[11pt]{article}
\usepackage{amsfonts}
\usepackage{graphicx,amscd,amsmath,amssymb,verbatim,amsthm,lscape,setspace}
\usepackage{dsfont}
\usepackage{float}
\usepackage{figsize}
\usepackage{rotating}
\usepackage{multirow}
\usepackage{caption}
\usepackage{booktabs}
\usepackage{bbm}
\usepackage{longtable}
\usepackage{color}
\usepackage{mathtools}
\usepackage{array}
\usepackage{threeparttable}
\usepackage[round]{natbib}
\usepackage{hyperref}
\usepackage{geometry}

\hypersetup{
  colorlinks   = true,   urlcolor     = black,   linkcolor    = black,   citecolor   = black }
\newcolumntype{K}[1]{>{\centering\let\newline\\\arraybackslash\hspace{0pt}}m{#1}}
\begin{document}

\title{Cooperation and Punishment Mechanisms in Uncertain and Dynamic Social
Networks\thanks{%
We are grateful to Marco Battaglini (editor), an anonymous associate editor, and two anonymous referees whose comments substantially improved the paper. We acknowledge the financial support from the Singapore Ministry
of Education Academic Research Fund (AcRF) Tier 2 grant (MOE2014-T2-1-105), City University of Hong Kong (CityU 11500017), and the Cambridge Endowment for Research in Finance (CERF). The paper benefited from discussions
during presentations at the 2019 North American ESA Meeting in Los Angeles
(USA) and the SEA 89th Annual Meetings in Fort Lauderdale (USA). \textit{%
Declarations of interest: none}. }}
\author{Edoardo Gallo\thanks{%
Faculty of Economics, University of Cambridge, Sidgwick Avenue, Cambridge
CB3 9DD, UK and Magdalene College, CB3 0AG, UK, phone: +44-1223-335286,
e-mail: edo@econ.cam.ac.uk.} \and Yohanes E. Riyanto\thanks{%
Division of Economics, School of Social Sciences, Nanyang Technological
University, 48 Nanyang Avenue, HSS 06-15A, Singapore 639818, phone:
+65-6592-1578, e-mail: yeriyanto@ntu.edu.sg.} \and Nilanjan Roy\thanks{%
Department of Economics and Finance, College of Business, City University of
Hong Kong, 83 Tat Chee Avenue, Kowloon Tong, Hong Kong, phone:
+852-3442-2659, e-mail: nilanroy@cityu.edu.hk.} \and Tat-How Teh\thanks{%
School of Management and Economics and Shenzhen Finance Institute, The
Chinese University of Hong Kong, Shenzhen, 2001 Longxiang Road, Longgang District, Shenzhen, China, e-mail: tehtathow@cuhk.edu.cn} }
\date{}
\maketitle

\begin{abstract}

This paper examines experimentally how reputational uncertainty and the rate of change of the social environment determine cooperation. Reputational uncertainty significantly decreases cooperation, while a fast-changing social environment only causes a second-order qualitative increase in cooperation. At the individual level, reputational uncertainty induces more leniency and forgiveness in imposing network punishment through the link proposal and removal processes, inhibiting the formation of cooperative clusters. However, this effect is significant only in the fast-changing environment and not in the slow-changing environment. A substitution pattern between network punishment and action punishment (retaliatory defection) explains this discrepancy across the two social environments.

\end{abstract}

{
}

\bigskip
\bigskip

\begin{small}
	
\end{small}

\newpage

\section{Introduction}

Understanding when and how repeated interactions lead to cooperation is
fundamental to comprehend many economic phenomena. Seminal theoretical
contributions have demonstrated that cooperative play is possible when
individuals are sufficiently patient
(e.g., \cite{friedman1971non}, \cite{fudenberg1986folk}), and several experimental
papers have shown that people cooperate if the gains to cooperation are
sufficiently large (e.g., \cite{dal2005cooperation}, \cite{dreber2008winners}).
Many of these studies, however, focus on a setup with two players
interacting with each other with perfect monitoring of actions, thus abstracting
away from some of the features present in real contexts where
cooperation operates, e.g., among several individuals connected through social networks.

A central feature of cooperative activities is that they often occur in a network of
relationships that change over time and consequently influence the diffusion of
behavior. Taking the network into consideration affects the effectiveness of the known mechanisms to sustain cooperation in non-network (bilateral) relationships and introduces novel ones. For
instance, action punishment by retaliating against a defector is a
well-documented channel to enforce cooperation in a bilateral setting. However, the presence of a network may render action punishment less effective
because a defection affects not only the original culprit but also other well-behaving (cooperative) partners. In the network context, instead, the removal of a link is a directly targeted and potentially more effective way to punish a defector.

The purpose of this paper is to investigate the emergence and evolution of
cooperation in a dynamic social environment. In our experiment, each subject is assigned to a group of 12 players who play a repeated Prisoner's Dilemma (PD) game. Subjects are connected
by a network in which a link (or a connection) denotes a pair of subjects who play a PD game
against each other in that round, and each subject is constrained to choose
either to cooperate or defect with all their neighbors. At the beginning of
a round, some subjects have the opportunity to remove some of their existing
links and/or propose new links to others. When making their decisions, subjects have access to information about the reputation of each of their connections in the form of their last five actions.

We adopt a $2 \times 2$ treatment design. The first dimension we
investigate is the rate of change of the social environment, defined as the
frequency with which opportunities to update one's links or form new links become
available (``\emph{Slow}'' and ``\emph{Fast}'' environments). The second dimension we investigate is the presence of uncertainty on whether each player's observed action corresponds to her intended
action (``\emph{No uncertainty}'' and ``\emph{Reputational uncertainty}''). These two aspects have
been investigated previously in isolation (e.g., \cite{fudenberg2012slow}, \cite{rand2011dynamic}). We extend the analysis of \cite{fudenberg2012slow} in a bilateral interaction setting to a more realistic social network setting.

The prototypical example of criminal activity to explain PD games is an ideal setting to illustrate the relevance of our design. Social networks are essential to conduct a large range of criminal endeavors (see, e.g., \cite{varese2011mafias}). In recent decades, the development of the internet has arguably made it easier for criminals to form and remove connections with partners in crime with the rise of various cybercrime platforms (\cite{lusthaus2018industry}). Our setup where links among individuals are evolving captures such a setting. Next, uncertainty also plays a prominent role in criminal activity in both offline and online settings, and the difficulty of monitoring makes it challenging to always learn the true intended actions from observed outcomes. In cybercrime markets, for instance, it is common for delivered services to be defective. For instance, stolen credit card numbers may not work because the seller has intentionally fabricated them or because of external factors such as the intervention of a law enforcement agency (e.g., \cite{sebagh2021exploring}). The reputational uncertainty element in our setting is meant to capture an environment with imperfect monitoring and uncertainty described in the example above.

Our first result is that reputational uncertainty has a first-order effect
on cooperation rate and welfare, while the rate of change of the social
environment has, at best, a second-order effect. Specifically, cooperation rate
and welfare significantly decrease with reputational uncertainty in both slow and fast environments. Meanwhile, they only qualitatively increase in a fast-changing social environment, holding the presence of uncertainty constant.

Even though the first result bears a resemblance to previous studies with two-player PD (see the survey at the end of this section), the reason why reputational uncertainty reduces cooperation in our setting is novel and distinct from these studies. In our dynamic social network setting, action choices of players co-evolve with the network structure, so that there are two levers for individuals to punish defectors and enhance cooperation:
\begin{itemize}
	\item \emph{Network punishment} deters defectors with the threat of having their links with cooperators removed;
	and/or cooperators avoiding to form new connections with defectors.
	\item \emph{Action punishment} is the standard threat of responding to a defector by engaging in retaliatory defection (which is also present in two-player PD games).
\end{itemize}
The use of network punishment has been documented in previous studies involving
endogenous updates to the interaction neighborhood without uncertainty in various types of network games (e.g., \cite{rand2011dynamic}, \cite{riedl2016efficient}). Compared to network punishment, action punishment is much less targeted in network settings because any retaliatory action/defection affects not only the defector but also the cooperative neighbors. This feature is absent in a two-player PD game because the consequence of an action punishment is felt solely by the target in such a setting. By exploring how the use of network and action punishments vary with our treatment variables, we can shed light on the impact of reputational uncertainty on cooperative behavior.

Our second set of results show that reputational uncertainty makes network punishment less prevalent -- in the sense that participants become more \emph{lenient} and \emph{forgiving} towards defectors. However, the magnitude of this effect differs across fast and slow environments. Our concepts of leniency and forgiveness extend the notions formulated by \cite{fudenberg2012slow} (for bilateral PD games) to a dynamic network setting involving more than two players. We show that, in the fast-changing social environment, reputational uncertainty makes players: (i) employ lenient strategies -- players are significantly less likely to terminate an existing relationship with someone having a history of low cooperative play
in an environment with uncertainty than without it; and (ii) employ forgiving strategies -- players are significantly more likely to propose a link to someone with a history of low cooperative play. While these effects of reputational uncertainty similarly occur in the slow-changing environment, they are marginally significant at best.

Why is it that reputational uncertainty affects the usage of network punishment differently across fast and slow environments? By design, reputational uncertainty means that a player's history of low cooperative play does not necessarily reflect the true intention of the player. As such, all else being equal, other players who are making network punishment decisions would be more willing to give the benefit of the doubt and to extend a second chance to this player with low cooperative history. In a fast environment, opportunities to use network punishment arrive frequently, so it is relatively inexpensive for players to be lenient and forgiving. In contrast, in a slow environment, being lenient and forgiving entails huge opportunity costs --- chances to use network punishment only arrive occasionally, so there is a risk of being forced to stay connected with a low-cooperativeness player for a long time. In broader contexts, our observation highlights that it is essential to consider the frequency with which punishment opportunities are available when researchers analyze the notions of leniency and forgiveness in punishment decisions.

A novel consequence of leniency and forgiveness in network punishment (as a result of reputational uncertainty) is that cooperative clusters fail to emerge, a feature absent in two-player games. Previous studies (e.g., \cite{rand2011dynamic}, \cite{wang2012cooperation}, \cite{gallo2015effects}, \cite{gallo_riyanto2019}) have shown that
reputational information and a fast-changing social environment promote cooperation by enabling the formation of clusters of cooperators who associate with each other and shun away defectors. We show that reputational uncertainty undermines this process because defectors are more likely to remain embedded in cooperators' neighborhoods, thereby creating an obstacle to the formation of cooperative clusters that would have driven up the cooperation level of the whole network.

Our final set of results shows that reputational uncertainty makes action punishment less prevalent, regardless of the updating rate of the social environment. Specifically, players' likelihood of switching from cooperation to defection becomes significantly less sensitive towards their neighbors' past defections when there is uncertainty. This weakens the extent of action punishment, which implies fewer repercussions from defecting, thus making defection more prevalent. Moreover, reputational uncertainty also lowers the cooperation rate through promoting opportunistic play -- players become more likely to defect after a history of highly cooperative play than in a setting without uncertainty. This is because players can hide behind the veil of uncertainty, which makes it difficult for others to infer whether the defection was intentional or not.

Combining our results on both network and action punishments highlights a novel interplay between these two dimensions in dynamic networks. In a slow-changing environment, players primarily rely on action punishment, given that chances to use network punishment are, by design, infrequent. As such, reputational uncertainty lowers cooperation primarily by dampening action punishment usage. In a fast-changing environment, players substitute action punishment with the more targeted network punishment so that reputational uncertainty reduces cooperation through its effect over both dimensions. To the best of our knowledge, such a substitution pattern between different punishment methods and its interaction with reputational uncertainty is new to the literature.

This paper contributes to a growing literature that sits at the intersection of, first, the vast body of work investigating the emergence and sustenance of cooperation using variations of the PD game, and, second, research on how networks affect behavior and economic activity. These literature span different disciplines, and it is not feasible to have an exhaustive review of all related contributions here. We aim, therefore, to position our paper in the context of the most relevant contributions with a bias toward the economic literature.

In experiments with an infinitely repeated two-player PD game under
imperfect public monitoring, \cite{aoyagi2009collusion} show that subjects
cooperate in an environment with a continuous public signal and that their
payoff decreases with the level of noise in the public signal. \cite%
{fudenberg2012slow} study a model of a repeated PD with reputational
uncertainty in players' actions and observe that, compared to the setting
with no uncertainty, individuals are slow to anger and fast to forgive.
Specifically, with uncertainty, subjects use strategies that, firstly, do not
prescribe reverting to punishment after a single bad signal, and secondly,
entail a return to cooperation after punishing the opponent. \cite%
{aoyagi2019impact} find similar behavior with private monitoring.\footnote{%
Early experimental studies have explored the effect of manipulating players'
information in repeated games in less standard ways. For example, \cite%
{cason1999laboratory} investigate a public goods game where information on
others' contributions are presented with some delay. \cite%
{bolton2005cooperation} study cooperation in an environment with limited
information about reputation. \cite{bereby2006speed} consider a repeated PD
game with noisy payoffs, even though in their experiment players could
monitor others' past actions perfectly.}$^{,}$\footnote{%
Related studies of repeated PD games under uncertainty include \cite%
{rojas2012role}, \cite{rand2015s}, \cite{embrey2013experimental}, and \cite%
{arechar2017m}.}

Beyond two-person games with uncertainty, \cite{ambrus2012imperfect} study
the effects of a costly punishment option on cooperation and social welfare
in long, finitely repeated three-person public good contribution games. They
show that the strong positive effects of increasing the severity of
punishment as identified in \cite{nikiforakis2008comparative} and \cite%
{egas2008economics} do not hold when there is uncertainty. With five-person
repeated public goods game with imperfect monitoring, \cite%
{ambrus2017democratic} further investigate the impact of democratic
punishment, when members of a group decide by majority voting whether to
inflict punishment on another member, relative to individual peer-to-peer
punishment.\footnote{See also \cite{deangelo2020peers} for a recent contribution on how endogenous monitoring structure interacts with punishment in the context of public good provision game. They also explore whether a subject's actions are automatically monitored and broadcasted to the whole group or whether the probability of monitoring is below 100\%. While we only consider the case where monitoring is automatic, the possibility of non-automatic detection offers an interesting avenue for future research.} Our contribution to this strand of work on cooperation in groups
in an uncertain reputational environment is to introduce the possibility for
individuals to choose whom to interact with given their (possibly imprecise)
information on others' reputations.

When there is perfect monitoring of others' actions, recent research has
demonstrated that the freedom to choose whom to interact with promotes cooperative behavior. \cite%
{rand2011dynamic} and \cite{wang2012cooperation} find that when subjects can
update their links frequently, cooperation is maintained at a high level
with participants preferentially removing links with defectors and forming
new links with cooperators. \cite{shirado2013quality} report that high
cooperation is achieved at an intermediate frequency of link updating.\footnote{%
When the frequency is too low, subjects choose to have many links, even if
they attach to defectors. When it is too high, cooperators cannot detach
from defectors as much as defectors re-attach and, hence, subjects resort to
behavioral reciprocity and switch their behavior to defection.} \cite%
{gallo2015effects} further show that in this context, knowledge about
everybody's reputation is crucial to achieve a high level of cooperation.%
\footnote{%
A substantial related literature studies how endogenous group formation
affects play in PD, public good and weak link games. Contributions include
\cite{ehrhart1999mobility}, \cite{hauk2001choice}, \cite{riedl2002exclusion}%
, \cite{coricelli2004partner}, \cite{cinyabuguma2005cooperation}, \cite%
{page2005voluntary}, \cite{guth2007leading}, \cite{ahn2009coming}, \cite%
{brandts2009competitive}, \cite{charness2014starting}, \cite%
{riedl2016efficient}, \cite{yang2017efficient}, \cite{riyanto2020highly}. See \cite{choi2016networks}
for a comprehensive review.} Our study contributes to this growing
literature by introducing reputational uncertainty to investigate how it
affects the effectiveness of the network formation mechanism in sustaining
cooperation.

It is possible to view the link formation and termination decisions in our networked setting as analogous
to the ability of individuals to employ targeted punishment in public goods games that allow for costly punishment. With regards to the role of punishment in inducing cooperation, starting
with \cite{fehr2000cooperation}, vast literature has demonstrated that the
possibility of costly punishment facilitates cooperation in public goods games. However, in the presence of noise, \cite%
{ambrus2012imperfect} show that costly punishment is less effective in
establishing cooperation. We investigate
whether the punishment mechanism through link formation/termination is less effective once we
introduce reputational uncertainty.

An alternative form of punishment that has been shown to promote
cooperation is ostracism %
(\cite{cinyabuguma2005cooperation}, \cite{maier2010ostracism}). Under ostracism,
individuals can expel each other from the group using various voting
schemes. The expelled players are then excluded from the benefits of any
cooperation undertaken by those still remaining in the group. We can think
of our set-up as allowing decentralized ostracism through the removal of
links by single individuals in an environment with reputational uncertainty.
Interestingly, we find that ostracism of defectors can be effective even with a
decentralized system, but only when there is no reputational uncertainty.

The remainder of the paper is organized as follows. Section 2 describes the design and procedures of the experiment. The results are presented in section 3, and the last section concludes.

\section{Experimental Design and Procedures}

In our experiment, each participant is assigned to a group of 12 who play a
game that is repeated for 25 rounds with random termination thereafter.\footnote{We opt for 25 rounds to ensure we have a long enough sequence of rounds to observe stationary play. The random termination protocol implemented after round 25 is to
eliminate end-game effects. An alternative design would be to either implement
a \emph{random termination protocol} in every round proposed by \cite{RothMurnighan1978} or \emph{a block random termination protocol} proposed by \cite{frechette2017infinitely}. The former is done by making a random draw with a pre-specified termination probability in every round to decide whether the round will be the terminal round. There are drawbacks of this method. First, we could potentially end up with a varying length of the replication rounds. Second, some of these replication rounds might be short while others might be long, making it difficult to observe stable patterns of network evolution. The block random termination protocol is a variant of the random termination protocol, in which participants interact in blocks of a pre-announced fixed number of rounds, for example, three blocks of 10 rounds. In each block, a terminal round is randomly drawn, but participants were not told which round it is. So they would treat all rounds within a block as payoff-relevant. Once the block is completed, participants are told whether there is a terminal round in the block and, if there is, which round it was. In hindsight, we could have adopted this protocol, which would overcome the drawbacks of the standard random termination protocol described earlier and ensure stationarity across all rounds of the experiment. However, we were not aware of this protocol when we ran our experiments.} At the beginning of a round, some participants have the chance to propose and/or remove links with some of the other group members. After the links have been updated, each participant plays a PD game with the other participants who are directly connected to her. We now describe the protocol of the game in detail, which consists of three stages in each round.

Figure \ref{design} illustrates the three stages that constitute a round. In
the initial round, all participants are always linked with each other, but in
any round thereafter (e.g., round T in Figure \ref{design}) they may not be.\footnote{In principle, any initial network structure that is not complete or not empty is arbitrary. Between these two contrasting configurations, a complete network has the advantage of allowing interaction between subjects to emerge more quickly, thus reducing the required length of experimental sessions.} In stage 1 of round T, $x$ random pairs of participants, i.e., those drawn with
dashed circles, may receive opportunities to propose and/or remove links
with each other. Suppose $i$ and $j$ is a random pair picked at time $T$,
each of them faces one of these choices:

\begin{figure}[tbph]
	\caption{\small Schema of the experimental design.}
	\vspace{-20pt}
	\begin{center}
		\includegraphics[width=0.92\textwidth]{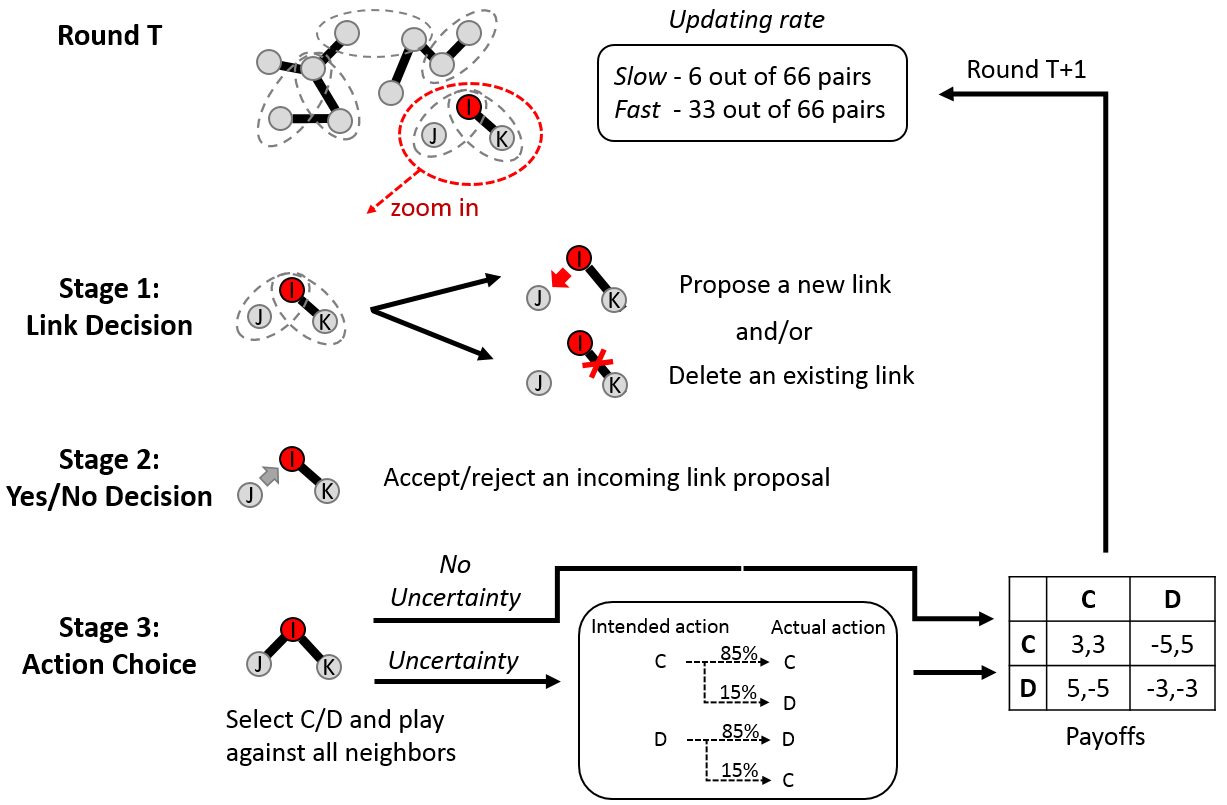} 
	\end{center}
	\label{design}
\end{figure}

\begin{itemize}
\item Remove the link if $i$ and $j$ were connected at round $T-1$. Deletion
of a link is unilateral and does not require mutual consent.\footnote{%
This is a natural assumption in the context of a relationship that is not
legally binding and is common in the economics literature %
(e.g., \cite{jackson1996strategic}).}

\item Propose a link if $i$ and $j$ were not connected at round $T-1$.
The formation of a link is bilateral and requires the consent of the recipient of
the proposal in Stage 2.
\end{itemize}

Stage 2 occurs if one and only one between $i$ and $j$ has proposed a link.
Given that link formation requires mutual consent, in stage 2, the recipient
of a link proposal has to decide whether to accept or reject it, as shown in
the third row of Figure \ref{design}. Notice that if two participants send a
link proposal to one another in Stage 1, the link is automatically
formed, and there is no prompt in Stage 2.

In stage 3, participants play the PD game shown at the bottom right of
Figure \ref{design} with each one of their neighbors, subject to the
restriction that they have to choose either action $C$ (cooperation) or
action $D$ (defection) for all neighbors.\footnote{%
We used neutral labels $A$ and $B$ in our experiment to avoid the
experimenter-demand effect. The selection of a payoff structure symmetric
around zero provides the correct incentives to allow the formation of
meaningful and realistic network structures because the symmetry of payoffs
in the gains/losses domains means that both the absence of a connection and
connections between a defector and a cooperator lead to no change in social
surplus, i.e., the sum of payoffs of all the participants. The only way to
produce a social surplus is a connection between two cooperators, and,
conversely, the only way to reduce the social surplus by an equal amount is a
connection between two defectors. This is in contrast to studies that have
non-negative or small negative payoffs, which lead to the emergence of
over-connected networks because the losses from being connected to a
defector are non-existent or negligible.} Choosing a single action
simultaneously toward all neighbors is a standard restriction employed in
the literature to allow for network effects, otherwise the network game
becomes a collection of independent PD games between pairs of players in the
group.\footnote{%
See for example \cite{morris2000network}, \cite{charness2014network}, \cite%
{riedl2016efficient}, and \cite{yang2017efficient}.} If a participant has no
neighbors, then she has no action to choose. Once all individuals make their choices, they see the points they receive
from the interaction with each of the other participants. They receive zero
points in this stage from participants with whom they are not connected to.
Subjects are also reminded of their chosen action and the action chosen by
their neighbors.

Appendix \ref{sec: instructions} shows screenshots for each of the three stages. For each
participant, the left area of the interface displays a visualization of
their links to their neighbors in which the nodes are colored in either
green or blue depending on whether their action in the last round was to
cooperate or defect respectively.\footnote{Note that participants have information about the local network structure (i.e., their neighbors), but they do not know the overall structure of the network. This is consistent with previous contributions in the literature, and it prevents information overload. Using a similar set-up, \cite{gallo2015effects} show that providing information about the whole network structure does not affect the aggregate level of cooperation, but it changes the distribution of cooperative activity.} The central part of the interface shows
the list of current neighbors with their past five actions and, at the end of
stage 3, the points obtained in the PD game with each of the neighbors.%
\footnote{\cite{cuesta2015reputation} show that cooperation is robust to
changing the number of past instances of play that participants see as long
as they can observe at least the last instance of play.} The right area of
the interface has the list of other participants who are not currently
neighbors. The bottom part prompts the participant to make the relevant
decision in each stage, e.g., the participants with whom she can
propose/remove a link in stage 1 and the choice of action in stage 3.

\textbf{End of experiment.} Each group of 12 participants plays at least 25 identical rounds of this
game. After round 25, there is a $50\%$ chance of termination to reduce any
endgame effects. Participants' earnings are based on six randomly selected
rounds. For each participant, two other participants are randomly picked for
each of the selected rounds, and the sum of the points obtained in the
interaction with these 12 randomly picked participants constitutes the total
sum of points earned by the participant in this part. Notice that it is
possible to pick an unconnected pair of participants and, in that case, the
earnings from that interaction are equal to zero.\footnote{%
This random selection of pairs for payment, independent of whether a
connection exists or not, ensures that there are uniform incentives
throughout the experiment in forming connections, so the payment system does
not introduce biases in the emerging network structure. For instance, if we
had excluded unconnected pairs from the random selection for payment, 
participants would have incentives to form just one link with a cooperator
to ensure that a specific pairing is picked. Some prior
studies (e.g., \cite{wang2012cooperation}, \cite{shirado2013quality}) pay the cumulative number of points participants have earned,
which may lead to satisficing in the later rounds and therefore lower
incentives to change the network.} Following the completion of the first part of the experiment, participants
take a \cite{holt2002risk} multiple price list risk-elicitation test and
fill in a socio-demographic questionnaire, which includes the standard
interpersonal trust question taken from the World Values Survey (WVS).%
\footnote{%
The interpersonal trust question is: \textit{``Generally speaking, would you
say that most people can be trusted or that you need to be very careful in
dealing with people?''.}}

\textbf{Treatment variations.} The boxes with round edges in Figure \ref{design} illustrate the two
treatment dimensions in the experiment, namely the \emph{updating rate} (%
\emph{slow} and \emph{fast}) of the social environment and \emph{whether
uncertainty is present}. We conduct four treatments and use a 2$\times $2
between-subject design to test the impact of uncertainty under two different
link formation/removal rates. Under \textit{no reputational uncertainty}
treatments, there is no uncertainty, so participants receive payoffs
according to their own action and neighbors' actions. Under \textit{%
reputational uncertainty} treatments, there is a commonly known 15\%
probability that an intended action is changed to the opposite action.\footnote{Previous experiments by \cite{fudenberg2012slow} and \cite{cason2019individual} select a slightly lower 12.5\% error rate.}
Subjects are informed of their intended action as well as their implemented
one. However, they are only notified of the neighbors' actual actions, not their intended ones.

The way we model reputational uncertainty
in our setup follows the paradigm proposed by \cite{fudenberg2012slow}. In
terms of the updating rate dimension, we consider a \textit{slow} $x=6$
condition in which only 9\% of pairings potentially change in each round,
and a \textit{fast} $x=33$ condition in which 50\% of pairings potentially
change in each round.

The data was gathered in 16 experimental sessions conducted at the Nanyang Technological University (NTU) in Singapore, and the experiment involved 384 undergraduate students from various majors. No subject participated in more than one session, and we used a between-subject design so that each subject participated in one and only one treatment. There are four sessions for each of the four treatments. Each session consists of 2 groups (networks) of size 12. This gives rise to a total of 8 independent observations at the network level per treatment. The choice of a group size of 12 ensures that the network is large enough to allow the emergence of interesting structural features while
at the same time it is small enough to make it feasible for us to collect a sufficient number of
independent observations per treatment in a laboratory setting. The experiment was programmed with
the software z-Tree \citep{fischbacher2007z}. Upon arrival, subjects were
seated at visually isolated computer workstations. Instructions were read
aloud, and subjects also received a copy of the instructions.\footnote{%
A sample copy of the instructions for the fast updating treatment with uncertainty is provided in
Appendix \ref{sec: instructions}.} Participants, who were each given a random user ID, remained completely anonymous to each other throughout the
experiment.

The sessions lasted approximately two hours, and participants earned on
average S\$13.73 in addition to the S\$2 show-up fee. Out of the 384
participants, 53\% are female, 62.4\% are Singaporeans, and the average age
is 21 years. About 23\% of the participants have had prior exposure to
game theory, and exactly half had participated in at least
one laboratory experiment previously. From the WVS interpersonal trust
question, 22\% believe that others can be trusted. In the post-experiment
questionnaire, almost all participants stated that they had no problem 
understanding the experiment.\footnote{%
The descriptive statistics of the participants in each of the four
treatments are provided in Appendix \ref{sec: demographics}.}

\section{Results}

\subsection{Overall cooperativeness}

\label{overall}

We start by analyzing how cooperation outcomes vary at the network level
across treatments, where there are eight independent observations for each
treatment. We adopt the following two network-level measures:

\begin{itemize}
\item \textbf{Cooperation rate. }The number of pairs with mutual (intended)
cooperation at the end of a round, divided by 66, which is the total number of
possible pairs in a complete network with 12 nodes.\footnote{%
Results remain the same if we measure cooperation based on the fraction of
intended cooperators, defined as the ratio of participants in a group who
choose $C$ as the intended action.}

\item \textbf{Welfare. }The total number of points obtained by all
participants from all interactions at the end of the round.
\end{itemize}

\begin{figure}[tbph]
\caption{\small Cooperation rate and welfare over time.}
\vspace{-20pt}
\begin{center}
\includegraphics[width=0.97\textwidth]{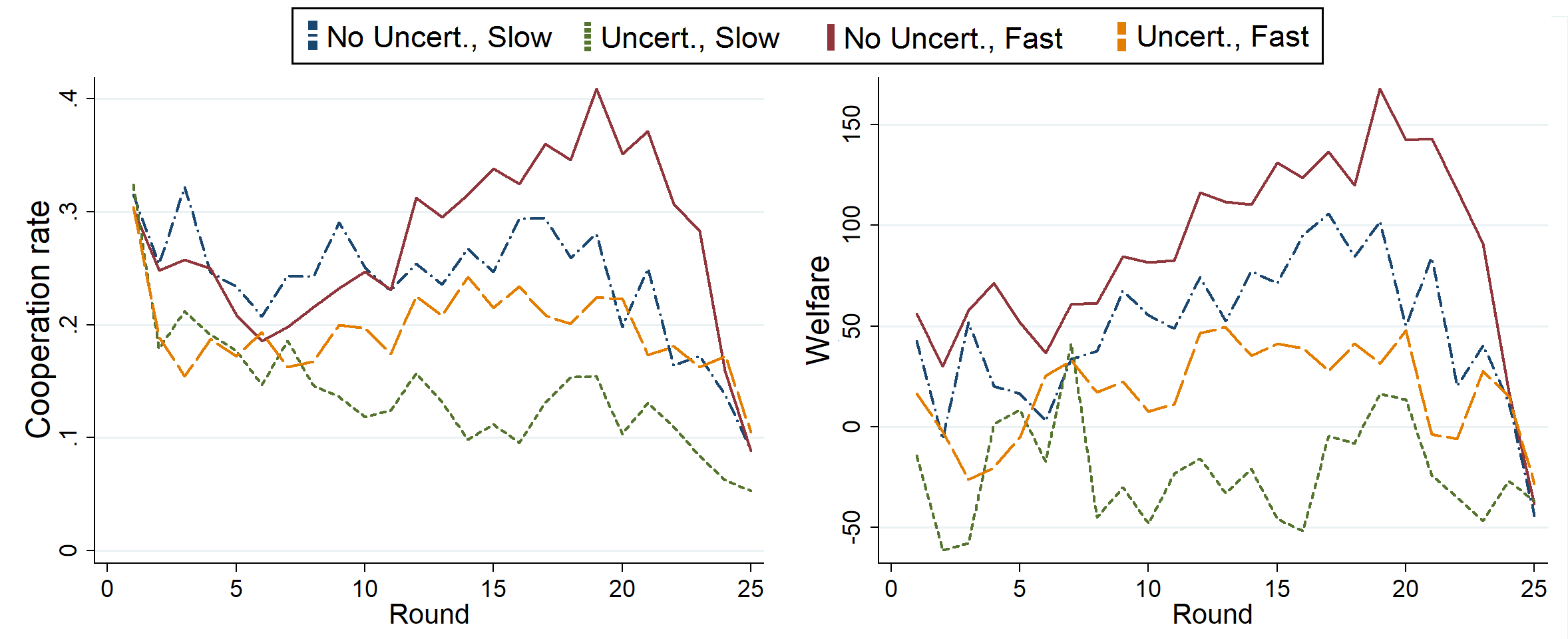} 
\end{center}
\label{fig:cooperation_series}
\end{figure}

Figure \ref{fig:cooperation_series} shows the evolution of cooperation rate
and welfare over time. After some initial volatility, cooperation rate and
welfare in different treatments tend to stabilize over time, aside from the end-game effects in the rounds toward the end of a session where we observe steep drops in both measures. To identify the treatment effects
formally, we utilize data from rounds 2-21 and exclude the last four rounds to avoid end-game effects, which are common in games of cooperation
(\cite{gachter2008long}, \cite{ambrus2012imperfect}, \cite{rand2011dynamic}, \cite{wang2012cooperation}).\footnote{We focus on rounds 2-21 throughout the paper. The only exception is analysis conditioned on reputation history, where we focus on rounds 6-21 because the first five rounds are needed to build a complete history of past plays.} Nonetheless, our results remain unchanged if we include the last four rounds as well by using data from rounds 2-25.

Table \ref{tab:coop welfare} provides summary statistics of the average cooperation rate and
welfare for each treatment, using the eight network-level observations per treatment. We apply the
non-parametric Kruskal-Wallis test to detect treatment effects, followed by
the Dunn's (\emph{DT}) test to examine the statistical significance of the
differences between any two treatments.\footnote{This methodology has been used in prior work involving network experiments in which the sample size at the network level is limited due to the number of participants required per network (e.g., \cite{gallo2015effects}). It is typical for network experiments to have 6-10 independent network-level observations in each treatment (\cite{rand2011dynamic}, \cite{riedl2016efficient}, \cite{yang2017efficient}, \cite{gallo_riyanto2019}).}

\begin{table}[h]
	\small
	\caption{\small Cooperation rate and welfare across treatments.}%
	\centering
	\begin{tabular}{lccccc}
		\hline\hline
		& \multicolumn{2}{c}{Slow} &  & \multicolumn{2}{c}{Fast} \\
		& No uncertainty & Uncertainty &  & No uncertainty & Uncertainty \\ \hline
		\\
  		Cooperation rate  \vspace{8pt} & 0.255 (0.063) & 0.144 (0.052) &  & 0.285 (0.080) & 0.198 (0.073) \\
		Welfare \vspace{8pt}           & 56.29 (74.93) & -20.29 (63.02) &  & 96.11 (66.11) & 20.96 (49.64) \\
		\hline\hline
	\end{tabular}%
	\begin{tablenotes}
		\footnotesize
		\item \qquad Note: we report mean values over rounds 2-21 with standard errors in parentheses.
	\end{tablenotes}
	\label{tab:coop welfare}
\end{table}

Holding the rate of network updating constant, the presence of reputational uncertainty
leads to significantly lower cooperation rate and welfare. In the \textit{slow} condition, reputational uncertainty reduces
cooperation rate from 0.255 to 0.144 ($p=0.003$, $DT$) and welfare
from 56.29 to -20.29 ($p=0.017$, $DT$). In the \textit{%
fast} condition, reputational uncertainty reduces cooperation rate from 0.285
to 0.198 ($p=0.024$, $DT$) and welfare from 96.11 to 20.96 ($p=0.028$, $%
DT$).

Meanwhile, holding the presence of reputational uncertainty constant, the introduction of a
fast-changing social environment increases cooperation rate and welfare, but the
effects are at best marginally significant. Without reputational uncertainty, the p-values ($DT$) for the difference between the \emph{slow} and \emph{fast} conditions are $p=0.345$ for cooperation rate and $p=0.182$ for welfare. With reputational uncertainty, the p-values for the difference between the \emph{slow} and \emph{fast} conditions are $p=0.071$ for cooperation rate and $p=0.132$ for welfare. We are, therefore, unable to replicate the strong finding from previous
studies on cooperation in dynamic network settings showing that a fast-changing
social environment increases cooperation %
(e.g., \cite{rand2011dynamic}, \cite{wang2012cooperation}). This is consistent with
the recent evidence showing that the role of network formation is not robust
to changes in the experimental environment (\cite{rand2014static}) suggesting
that its role is second order with respect to reputation.

Finally, note that the effects we find are not driven by initial conditions.
There is no significant difference in the cooperation rate between any two
treatments in round 1 ($p>0.1$ for all comparisons, $DT$). Because in all
treatments the groups start, on average, from the same level of cooperation
rate, it must be that reputational uncertainty and the dynamics of network
evolution are the factors that drive the differences in the cooperation rate
that emerge over time.

Our first result is, therefore, the following.

\bigskip

\textbf{Result 1}: (i) \textit{Reputational uncertainty significantly
reduces cooperation rate and welfare in both slow- and fast-changing conditions. (ii)
The fast-changing social environment increases cooperation rate
and welfare, but only qualitatively, regardless of the presence of
reputational uncertainty.}

\bigskip

The significant finding in the first part of Result 1 is consistent with the
insights from the existing studies on two-person cooperation games and
public good games without any network structure. These studies have shown
that the presence of reputational uncertainty (defined in a broad sense) results in
strictly lower cooperation (\cite{aoyagi2009collusion}, \cite{aoyagi2019impact}) and lower contributions (\cite{ambrus2012imperfect}, \cite{ambrus2017democratic}). However, in our dynamic social network setting, the reason why reputational
uncertainty reduces cooperation is different from these studies.  In the subsequent sections, we explore two individual-level mechanisms driving Result 1, which we briefly outline below.

In Section \ref{social}, we show that reputational uncertainty makes subjects more
\emph{lenient} and \emph{forgiving} (to be made precise below) in their link
removal/formation decisions, i.e., it lowers the use of network
punishment (Results 2 and 3 below). Cooperators are unable to precisely break links with defectors
and form links with other cooperators (Result 4 below). Consequently, reputational
uncertainty disrupts the so-called \textquotedblleft assortative matching
effect\textquotedblright\ of network dynamics, whereby cooperators cluster
among themselves and shun away defectors to sustain cooperation %
(e.g., \cite{rand2011dynamic}, \cite{wang2012cooperation}, \cite{gallo2015effects}, \cite{gallo_riyanto2019})%
.\footnote{See also \cite{riedl2016efficient}, \cite{yang2017efficient}, and \cite{riyanto2020highly} for the same mechanism in the context of dynamic network coordination games.} While this mechanism of reputational uncertainty is present in both \emph{slow} and \emph{fast} conditions, it is
statistically significant only in the fast condition. We explain this difference by pointing out a substitution pattern between network and action punishments across \emph{slow} and \emph{fast} conditions (Result 6 below).

In Section \ref{action}, we show that reputational uncertainty makes subjects' choice of action (to cooperate or to defect) less sensitive toward past defections by other players. In other words, action punishment is less prevalent in the presence of reputational uncertainty, in both \emph{slow} and \emph{fast} conditions (Result 5 below). This reduction in punishment, in turn, promotes defections and opportunistic plays with reputational uncertainty (Result 7 below).

\subsection{Network punishment}

\label{social}

\emph{Network punishment} is a primary tool to maintain cooperation when
players can choose their interaction neighborhood. It allows cooperators
to remove links with defectors in order to shun them away while at the same time
selectively form connections with other cooperators. We distinguish between link
removal and link formation decisions and define the following two notions
that describe subjects' strategies in these decisions.

\begin{itemize}
\item \textbf{Leniency. }Strategies are lenient if, conditional on receiving
an opportunity to remove a link, the deciding player does not remove the
existing link with a neighbor who has a history of low cooperative play (i.e.,
whose actual action was $C$ at most once in the last five rounds).

\item \textbf{Forgiveness.} Strategies are forgiving if, conditional on
receiving an opportunity to propose a link, the deciding player proposes a
link to someone with a history of low cooperative play (i.e., whose actual
action was $C$ at most once in the last five rounds).
\end{itemize}

These notions of leniency and forgiveness are novel as they are designed
to capture variations in link formation/removal rather than action
punishment which is the focus in two-person PD games %
(\cite{fudenberg2012slow}, \cite{embrey2013experimental}, \cite{aoyagi2019impact}).\footnote{%
In these studies, strategies are defined to
be lenient if they do not prescribe sure defection following a single
defection by the opponent, and forgiving if they return to cooperation after
having defected. The main finding is that in the presence of reputational
uncertainty subjects use strategies that are both lenient and forgiving.} We
condition our definitions on play histories of length-5 because in our
experiment the computer interface shows each subject the list of her current
neighbors with their past five actions.

It should be emphasized that the exact definitions of leniency and
forgiveness above are empirically motivated for our analysis.
Our results remain robust to other definitions. For example, we can alternatively define
\textquotedblleft a history of low cooperative play\textquotedblright\ as
actual action being $C$ at most two or three times in the last five rounds. Furthermore, we can broaden the definition of forgiveness by including link acceptance decisions as well. See Appendices \ref{sec: alt definition} and \ref{sec: accept} for further details.

In what follows, we focus on identifying the effect of reputational
uncertainty on leniency and forgiveness in network punishment.\footnote{%
For completeness, in Appendix \ref{sec: dynamic effect}, we report
the effect of the network updating rate on leniency and forgiveness.} Given the
definitions of leniency and forgiveness, throughout this section, we focus
on observations from rounds 6-21, unless otherwise stated. Before
proceeding further, we first describe the several standard network metrics 
prominently used in the analysis.\footnote{See \cite{jackson2008} for more details.}


\begin{itemize}
\item \textbf{Degree}. The degree $n_{i}$ measures the number of links or
neighbors that a player $i$ has.

\item \textbf{Clustering coefficient}. The local clustering coefficient
measures the connectedness of a player's neighbors. A player $i$ with $n_{i}$
neighbor has $\binom{n_{i}}{2}$ possible pairwise links between her
neighbors. The clustering coefficient measures the ratio of the number of
links that player $i$'s neighbors have to the total number of possible links
that can potentially be formed among player $i$'s neighbors.

\item \textbf{Betweenness centrality.}\footnote{%
Our main insights remain robust if we use closeness centrality instead.}\textbf{\ }The
betweenness centrality $c_{B}\left( i\right) $ measures the importance of a
player $i$ in a network in terms of the flow in the network.
Formally,
\begin{equation*}
c_{B}\left( i\right) =\sum_{j\neq k\neq i}\frac{\sigma _{jk}(i)}{\sigma _{jk}%
},
\end{equation*}%
where $\sigma _{jk}$ is the number of shortest paths between players $j$ and
$k$, and $\sigma _{jk}(i)$ is the number of those shortest paths that go
through player $i$.

\end{itemize}

\subsubsection{Reputational uncertainty promotes leniency and forgiveness}

\label{leniency and forgiveness}

\textbf{Analyzing Leniency}. We begin by analyzing link removal decisions. Figure \ref%
{fig:historemove2} displays the average likelihood of removing an existing
link, conditional on receiving an opportunity to do so (recall that such an
opportunity arrives at an exogenous rate). We consider three different
histories of a neighbor's cooperative play: 0 or 1 $C$ (\textit{low}), 2 or
3 $C$ (\textit{medium}) and 4 or 5 $C$ (\textit{high}). From Figure \ref%
{fig:historemove2}, we can see that subjects seldom remove existing links that are highly cooperative (around 2\% of
the total observations in each treatment), with or without reputational
uncertainty. In contrast, in instances of low or medium cooperative play of
the existing link, a subject is less likely to remove a link when there is
reputational uncertainty, suggesting that the subjects become more lenient.
\medskip
\begin{figure}[tbph]
\caption{\small Average probability to remove an existing link, conditional on
receiving an opportunity.}
\vspace{-20pt}
\begin{center}
\includegraphics[width=0.9\textwidth]{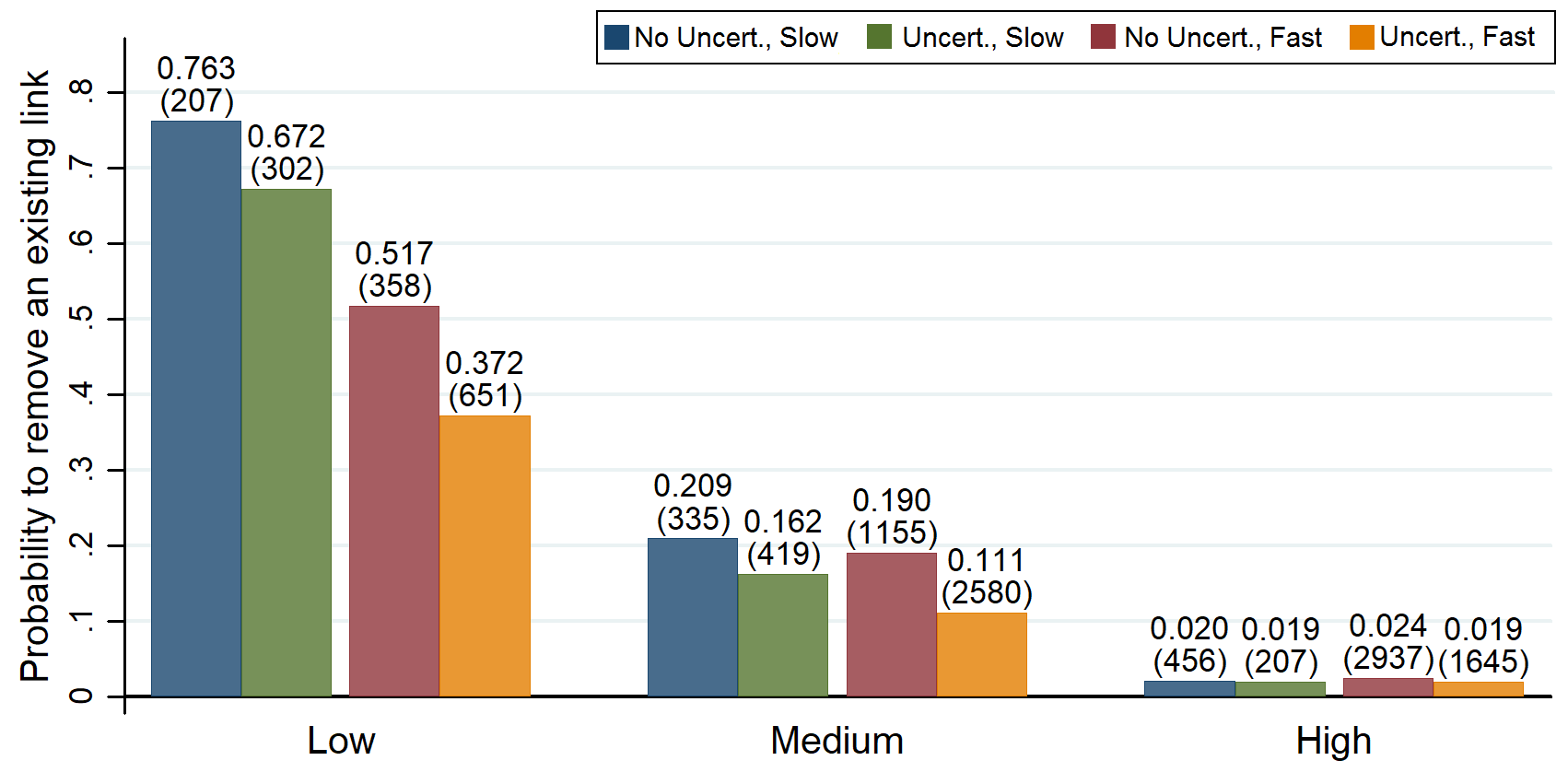}
\end{center}
\vspace{-20pt}
\caption*{\footnotesize Notes: Separated by the last five rounds' history, where \textit{Low}: the actual action was $C$ at most once in the last five rounds;
\textit{Medium}: $C$ was the actual action two or three times; \textit{High}%
: $C$ was the actual action more than three times. We report the average values over rounds 6-21, with the number of
observations in parentheses.}
\label{fig:historemove2}
\end{figure}

To test the effect of reputational uncertainty on leniency, we report in
Table \ref{tab:lenient} the results from a random-effects logit regression
with the dependent variable being the link removal decision ($=1$ if the
link is unilaterally removed). We conduct the regressions separately for the
\emph{slow} and \emph{fast} conditions, and for neighbors with histories of
\emph{low} and \emph{medium} cooperative plays. We exclude link removal
decisions with neighbors with histories of \emph{high} cooperative play to simplify
the presentation, especially given that subjects virtually never remove
links with such neighbors, as observed from Figure \ref{fig:historemove2}.
We cluster the standard errors at the group level and use observations from
rounds 6-21.

The independent variables include the following: a treatment
dummy variable for \emph{Reputational uncertainty} ($=1$ if there is
uncertainty) and network metrics of the deciding player (\emph{Degree}, \emph{Local} \emph{clustering}, \emph{%
Centrality}). To control for the recent cooperativeness of the deciding player, we also include the last intended action chosen by the deciding player (\emph{Own C Action$_{t-1}$}), which equals $1$ if it is $C$ and $0$ if it is $D$.\footnote{Results 2 and 3 below remain valid if we replace this control variable with the deciding player's ``innate cooperativeness'' defined in Section 3.2.2.} In addition to these variables, we also include
seven other control variables: round number, age, gender, trust obtained from the World Values Survey question,
risk profile elicited from the Holt-Laury procedure, and prior exposure to game
theory concepts and incentivized economic experiments.\footnote{%
All regression results reported in the main text remain robust if we exclude network metrics and these seven control variables. See Appendix \ref{sec: alt definition} for
further details.}

\begin{table}[h]
	\small
	\caption{\small Random-effects logit regressions - effect of reputational uncertainty on leniency.}%
	\centering
	\begin{tabular}[c]{lccccc}
		\hline\hline
		 &\multicolumn{2}{c}{\textit{Slow}}    &   \qquad\qquad\qquad   & \multicolumn{2}{c}{\textit{Fast}} \\
		History of the target neighbor: \qquad & {Low $C$}   & {Medium $C$}  & &  {Low $C$}  & {Medium $C$}      \\
		\hline		
		\multicolumn{6}{l}{\textbf{Dependent variable: Link removal} (conditional on receiving an opportunity)}\\
		&&&&&\\
		Reputational uncertainty & -0.397   & -0.372   &  & -0.949** & -0.820*** \\
		& (0.394)  & (0.323)  &  & (0.375)  & (0.207)   \\
		Own C Action$_{t-1}$       & 1.380*** & 0.703*** &  & 0.360**  & 0.404     \\
		(recent cooperativeness)& (0.381)  & (0.153)  &  & (0.181)  & (0.253)   \\
		&          &          &  &          &           \\
		Degree                   & 0.193    & -0.219** &  & 0.014    & -0.068    \\
		& (0.155)  & (0.091)  &  & (0.051)  & (0.044)   \\
		Local clustering         & 1.223    & 1.971**  &  & -0.401   & 0.589     \\
		& (1.211)  & (0.978)  &  & (0.346)  & (0.468)   \\
		Betweenness centrality   & -4.723   & 5.967    &  & -3.842** & -0.938    \\
		& (6.785)  & (4.905)  &  & (1.842)  & (1.538)   \\
		&          &          &  &          &         \\ 	\hline
		Control and intercept & Yes & Yes    &         & Yes & Yes              \\
		Observations          & 508    &  754   &  & 1009 & 3735           \\
		\hline\hline
	\end{tabular}
\begin{tablenotes}
	\footnotesize
	\item Notes: \textit{Low $C$}- $C$ was the actual action at most one time in last 5 rounds; \textit{Medium $C$}- $C$ was the actual action two or three times. Standard errors (clustered at the group level) are in parentheses, {*} \(p<0.10\), {**} \(p<0.05\), {***} \(p<0.01\). Control variables include round number, age, gender, trust and risk aversion profiles, and prior exposure to game theory and economic experiments.
\end{tablenotes}
	\label{tab:lenient}
\end{table}

The main insight from Table \ref{tab:lenient} comes from the coefficient of
\emph{Reputational uncertainty}. It shows that reputational uncertainty
results in subjects using strategies that are more lenient in the \emph{fast}
condition, in the sense that they become significantly less likely to remove
existing links with neighbors of low and medium cooperative play ($p<0.05$
and $p<0.01$, respectively). In contrast, in the \emph{slow} condition,
reputational uncertainty does not significantly affect leniency, given that
the presence of uncertainty does not significantly affect the probability of link
removal for any history ($p>0.1$). In addition, such removal decisions are
primarily triggered by deciding players whose action is $C$ in the previous
round (coefficient of \emph{Own C Action$_{t-1},$ }$p<0.05$), confirming the
intuition that players who are ``recently'' cooperative (prior to the link removal decision) punish defectors (\textit{Low $C$}) by removing links with them.

\bigskip
\textbf{Result 2}: \textit{Reputational uncertainty increases leniency in
link removal decisions, but the effect is statistically significant only in
the fast condition.}

\bigskip

In Appendix \ref{sec: leniency and linknature}, we show that Result 2 remains robust even if we take into account that subjects who are making link removal decisions may treat links that were formed from their own proposal differently compared to links they accepted but did not initially propose. Interestingly, we find that subjects are indeed significantly less lenient toward the latter type of links (but only in the fast condition).\footnote{We thank an anonymous referee for suggesting this exercise.} Thus, leniency depends on how a relationship was initially formed, a novel feature that is absent in non-network settings.

\bigskip

\begin{figure}[h]
\caption{\small Average probability to send a proposal, conditional on
	receiving an opportunity.}
\vspace{-20pt}
\begin{center}
\includegraphics[width=0.9\textwidth]{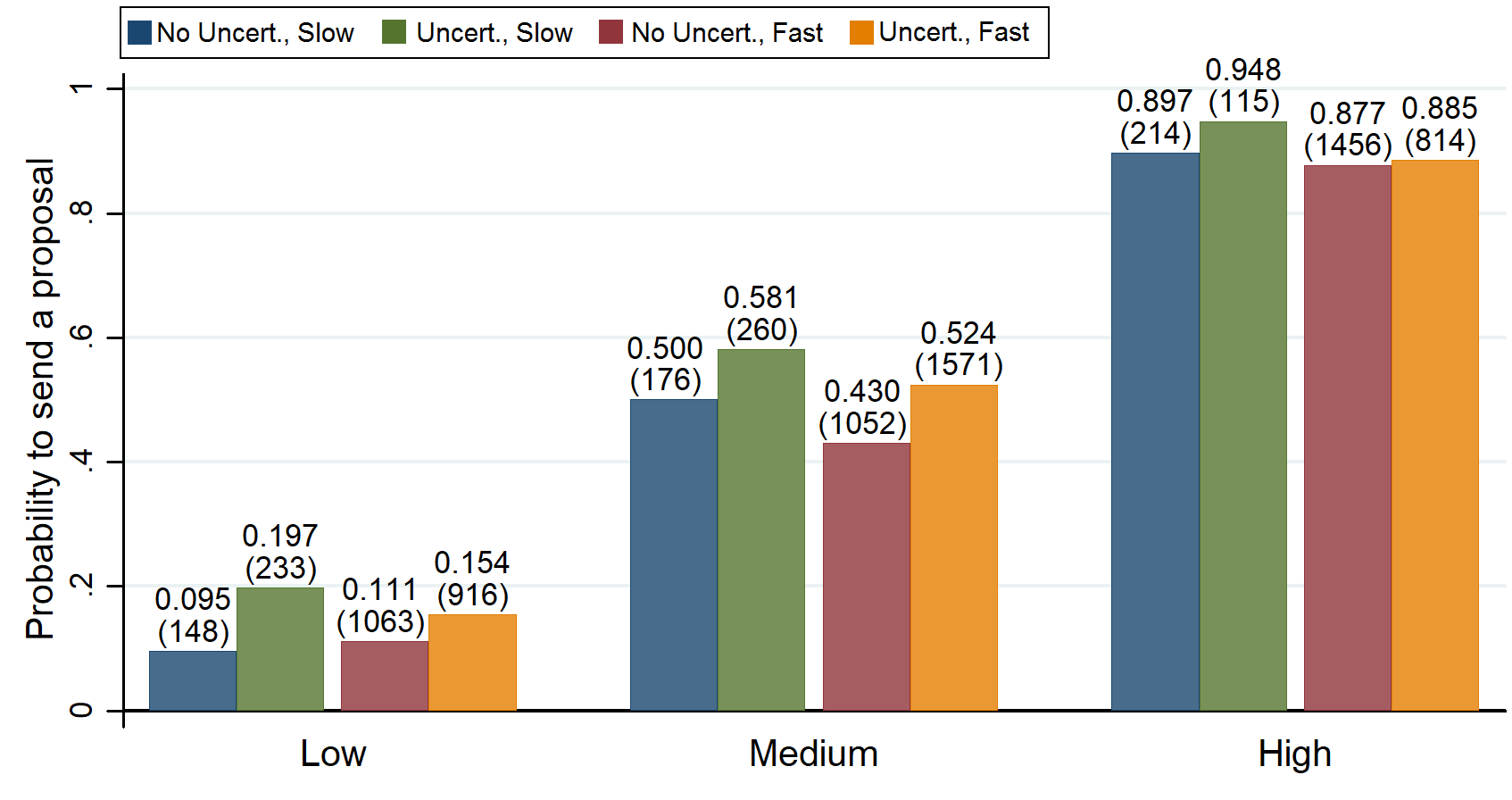}
\end{center}
\vspace{-20pt}
\caption*{\footnotesize Notes: Separated by the last five rounds' history, where \textit{Low}: the actual action was $C$ at most once in the last five rounds;
	\textit{Medium}: $C$ was the actual action two or three times; \textit{High}
	: $C$ was the actual action more than three times. We report the average values over rounds 6-21, with the number of
	observations in parentheses.}
	\label{fig:forgiving}
\end{figure}

\textbf{Analyzing Forgiveness}. Next, we analyze forgiveness in link formation decisions. Figure \ref
{fig:forgiving} displays the average probability of sending a proposal
(conditional on receiving an opportunity to do so) for the three different
histories of the potential neighbor's cooperative play defined earlier. We
observe that the likelihood of sending a proposal is consistently high
(close to 0.90) for recipients with a history of high cooperative play, with
or without reputational uncertainty. In contrast, this likelihood is
generally less than 0.58 when a potential neighbor has a history of low or
medium cooperative play. However, reputational uncertainty results in
subjects being more willing to send a proposal to such potential neighbors,
suggesting that the subjects become more forgiving.

\medskip

To test the effect of reputational uncertainty on forgiveness, we report in
Table \ref{tab:forgiving} the results from a random-effects logit regression
with the dependent variable being the link proposal decision ($=1$ if the
participant sends a proposal). Our regression specifications and control variables used in the regressions shown in Table
\ref{tab:forgiving} are the same as those in Table \ref{tab:lenient}. Again, the main insight from Table \ref{tab:forgiving} comes from the
coefficient of \emph{Reputational uncertainty}. In the \textit{fast} condition, reputational uncertainty
makes the subjects more forgiving when forming new links. They become
significantly more likely to propose new links with target potential neighbors of low and
medium cooperative play ($p<0.05$ and $p<0.01$, respectively).\footnote{This result remains robust even if we additionally control for the number of links of the target potential neighbor.} In contrast,
in the \emph{slow} condition, reputational uncertainty does not
significantly affect forgiveness given that the coefficient of \emph{
Reputational uncertainty }is insignificant for any history ($p>0.1$).
Finally, consistent with the intuition discussed at the beginning of Section
\ref{social}, recent cooperative players are less likely to form new links with
neighbors of low and medium cooperative play (coefficient of \emph{Own C Action
$_{t-1},$ }$p<0.05$).

\medskip
\begin{table}[h]
	\small
	\caption{\small Random-effects logit regressions - effect of reputational uncertainty on forgiveness.}
	\centering
	\begin{tabular}[c]{lccccc}
		\hline\hline
		&\multicolumn{2}{c}{\textit{Slow}}    &   \qquad\qquad\qquad   & \multicolumn{2}{c}{\textit{Fast}}             \\
		History of the potential neighbor: \qquad & {Low $C$}         & {Medium $C$}  & &  {Low $C$}    & {Medium $C$}      \\
		\hline
		\multicolumn{6}{l}{\textbf{Dependent variable: Link proposal} (conditional on receiving an opportunity)}\\
		&&&&&\\
		Reputational uncertainty & 1.684   & 0.145     &  & 0.789**   & 0.763***  \\
		& (1.760) & (0.341)   &  & (0.393)   & (0.205)   \\
		Own C Action$_{t-1}$       & -1.043  & -1.077*** &  & -0.424**  & -0.402*** \\
		(recent cooperativeness)& (0.679) & (0.251)   &  & (0.178)   & (0.154)   \\
		&&&&&\\
		Degree                   & -0.294  & -0.174    &  & -0.189**  & -0.141*** \\
		& (0.272) & (0.124)   &  & (0.074)   & (0.046)   \\
		Local clustering         & -3.809* & -0.609    &  & -1.044*** & -0.244    \\
		& (2.160) & (1.116)   &  & (0.317)   & (0.310)   \\
		Betweenness centrality   & 14.463* & 2.401     &  & 2.855     & 4.984**   \\
		& (8.452) & (8.932)   &  & (2.051)   & (2.023)  \\
		&&&&&\\
		Control and intercept & Yes & Yes    &         & Yes & Yes              \\
		Observations          & 381    &  436   &  & 1979 & 2623           \\
		\hline\hline
	\end{tabular}
	\begin{tablenotes}
		\footnotesize
		\item Notes: \textit{Low $C$}- $C$ was the actual action at most one time in last 5 rounds; \textit{Medium $C$}- $C$ was the actual action two or three times. Standard errors (clustered at the group level) are in parentheses, {*} \(p<0.10\), {**} \(p<0.05\), {***} \(p<0.01\). Control variables include round number, age, gender, trust and risk aversion profiles, and prior exposure to game theory and economic experiments.
	\end{tablenotes}
	\label{tab:forgiving}
\end{table}

\bigskip
\textbf{Result 3}: \textit{Reputational uncertainty increases forgiveness in
link proposal decisions, but the effect is statistically significant only in
the fast condition.}

\bigskip

From Results 2 and 3, we note that reputational uncertainty affects the usage of network punishment differently across the \emph{slow} and \emph{fast} conditions. Intuitively, when there is reputational uncertainty, a deciding
player finds it difficult to ascertain whether a specific \emph{target player} with a
history of low/medium cooperative is a \textquotedblleft true
defector\textquotedblright . The observed history of low/medium cooperative
play could be a result of unfortunate random draws that reverse the target
player's intention to cooperate. The standard intuition from two-person PD
games (\cite{fudenberg2012slow}, \cite{embrey2013experimental}, \cite{aoyagi2019impact})
suggests that, facing such uncertainty, the deciding player would give
the target player a \textquotedblleft benefit of doubt\textquotedblright,
and adopt strategies that are more lenient and forgiving. This is precisely
the intuition in the \emph{fast} condition in Results 2 and 3.

In contrast, in the \emph{slow} condition, there is a high potential burden for
maintaining/starting a relationship with a target player having a history of low
cooperative play. If the target player turns out to be a genuine defector, the
next opportunity for the deciding player to remove this link would only
arise in the distant future. In other words, in the \emph{slow} condition,
giving a \textquotedblleft benefit of doubt\textquotedblright\ is costly,
thus explaining why the effects of reputational uncertainty are not
statistically significant.\footnote{To complement this argument, we show in Appendix \ref{sec: dynamic effect} that subjects indeed become significantly more lenient in the fast condition than in the slow condition, holding the presence of reputational uncertainty constant.}

\subsubsection{Reputational uncertainty disrupts clustering of cooperators}

\label{network}

In this section, we examine how the increase in subjects' leniency and forgiveness in network punishment (as a consequence of the reputational uncertainty) affects the network-level metrics and the tendency for cooperators to connect to each other.

We classify a subject as an \emph{innate cooperator} (Type-$C$)
if she chose $C$ as her first-round action, and an \emph{innate defector} (Type-$D$)
if she chose $D$. This categorization is suitable for our homophily analysis below because it is exogenous
to the evolution of cooperation in the group so that it is fixed before
participants receive any feedback about game play. The average number of
innate cooperators ranges from 6.63 to 6.88 across treatments, and there are no
significant differences across treatments ($p>0.1$, $DT$). Appendix \ref{sec: type} provides further details on the comparison between these two
categories of players and the statistical analyses. Most importantly, as one would expect if our
categorization into these types is meaningful, we observe that Type-$C$ players have significantly more intended cooperative actions compared to Type-$D$ players ($p<0.05$).

Following \cite{currarini2009economic}, we measure the tendency for
players of the same type $i$ to connect to each
other as follows. Consider a population with two categories of players $i\in
\left\{ C,D\right\} $. Let ${w_{i}}$ be the relative fraction of $i$ players
in the population, where ${w_{C}+w}_{D}=1$. Let ${s_{i}}$ denote the average
number of links that individuals of category $i$ have with players who are
of the same category and let ${o_{i}}$ be the average number of links that
category $i$ players have with players of the opposite category. The
homophily index is
\begin{equation*}
{H_{i}=}\frac{{s_{i}}}{{s_{i}+o_{i}}},
\end{equation*}
and it captures the fraction of the links of players of category $i$ that
are with that same category. A key drawback of the homophily index is that
it is sensitive to the relative size of the two categories in the
population, so that it is \emph{not a suitable measure} to evaluate the
extent of homophily between categories of different sizes (e.g., in our
setting). For this reason, we instead focus on the following measure in our
analysis:

\begin{itemize}
\item \textbf{Inbreeding Homophily index (IH). }Inbreeding homophily index
of category $i$ is defined as
\begin{equation*}
{IH_{i}=}\frac{{H_{i}-w_{i}}}{{1-w_{i}}}\text{{.}}
\end{equation*}
Intuitively, it corrects the homophily index by the relative sizes of each
category. We say that there is inbreeding homophily between individuals of
category $i$ if ${IH_{i}>0}$, and there is baseline homophily if ${IH_{i}=0}$
.
\end{itemize}

Table \ref{tab:network stat} provides summary statistics of the standard
network-level metrics, $IH$ of type-$C$ players, and the average betweenness centrality
of each type-$C$ player. We focus on rounds 2-21 (as in Table \ref{tab:coop welfare}) because
we do not need to condition the analysis on play histories of length-5
unlike in the previous subsection. We then apply the Mann-Whitney rank-sum
test (\emph{MW}) to evaluate the effect of reputational uncertainty, holding constant
the rate of network updating.

\medskip
\begin{table}[h]
	\small
	\caption{\small Network-level metrics across treatments.}
	\centering
	\begin{tabular}{lccccc}
		\hline\hline
		& \multicolumn{2}{c}{Slow} &  & \multicolumn{2}{c}{Fast} \\
		& No uncertainty & Uncertainty &  & No uncertainty & Uncertainty \\ \hline
		\\
		Avg. degree  \vspace{8pt}   					& 7.485 (1.537) & 7.293 (1.621)  && 6.184 (1.316) & 6.729 (1.150)  \\
		Avg. local clustering    \vspace{8pt}   		& 0.710 (0.149) & 0.681 (0.156)  && 0.675 (0.113) & 0.712 (0.107)  \\
		IH of type-$C$ players    \vspace{8pt}  	 	& 0.061 (0.064) & -0.024 (0.126) && 0.122 (0.137) & -0.006 (0.104) \\
		Avg. btw. centrality of type-$C$ players \vspace{8pt} 	& 0.038 (0.009) & 0.038 (0.009)  && 0.051 (0.010) & 0.041 (0.011) \\
		\hline\hline
	\end{tabular}%
	\begin{tablenotes}
		\footnotesize
		\item Note: we report mean values over rounds 2-21 with standard errors in parentheses.
	\end{tablenotes}
	\label{tab:network stat}
\end{table}

The main finding from Table \ref{tab:network stat} is that, in the fast condition,
reputational uncertainty significantly (i) decreases the
inbreeding homophily index ($IH$) of type-$C$ players from 0.122 to -0.006 ($MW$,
$p=0.046$); and (iii) decreases the average betweenness centrality of each type-$C$ player from 0.051 to 0.041 ($MW$, $p=0.059$). However, the differences in these measures are statistically insignificant in the slow condition ($MW$, $p>0.1$). Finally, reputational
uncertainty has no significant effect on average degree and local clustering
in both fast and slow conditions.\footnote{We also perform Kolmogorov-Smirnov equality test to check the effect of reputational uncertainty on the overall degree distributions. We find no significant effect ($p>0.1$), conditioned on either fast or slow treatment.}

\bigskip

\textbf{Result 4}: \textit{Reputational uncertainty decreases (i) inbreeding
homophily among innate cooperators and (ii) average centrality of each innate cooperator, but only in the fast condition.}

\bigskip

Result 4 reflects the intuition that innate cooperators may find it more difficult
to recognize each other and connect when there is reputational uncertainty. This
results in a lower level of cooperativeness-based homophily and a lower centrality of each innate cooperator, especially when the network updating happens quickly.

\subsection{Action punishment and opportunistic play}

\label{action}
Action punishment is the primary channel to ensure cooperative behavior in
two-player PD games. It entails defecting as a response to defection by
the other player. The following definition extends this notion to our set-up
with more than two players in a linked network.

\begin{itemize}
\item \textbf{Action Punishment. }The likelihood of a player choosing $D$ as the
intended action as an increasing function of the fraction of neighbors who choose $D$. A steep reaction function indicates a strong extent of action punishment and vice versa.
\end{itemize}

\subsubsection{Reputational uncertainty erodes action punishment}

Intuitively, the presence of reputational uncertainty could affect a
player's choice of action in two ways. First, a player finds it difficult to
ascertain whether defection choices by her neighbors reflect their true
intentions or are merely random noise
(\cite{fudenberg2012slow}, \cite{embrey2013experimental}, \cite{aoyagi2019impact}). Thus, with uncertainty, the
player's likelihood of choosing $D$ would be less sensitive towards her
neighbors' past defections, weakening the extent of action punishment. Second, the weaker action punishment implies less repercussions from
defecting, making defection more tempting. Consequently, holding constant the neighbors' behaviors, we expect a player's likelihood of choosing $D$ to be higher when there is uncertainty. In sum, if we plot a subject's likelihood of choosing $D$ as a
function of the fraction of neighbors who choose $D$, reputational
uncertainty should raise the intercept while flattening the gradient of the
plot.

We report in Table \ref{tab:action} the results from a random-effects logit
regression with the dependent variable being each subject's choice of
intended defecting action ($=1$ if it is $D$) in each round $t$. We focus on the switching behaviors by limiting the analysis to the subsample of instances when a subject has selected $C$ as the
intended action in round $t-1$ and focus on rounds 2-21.\footnote{The results remain the same if we do not limit the analysis to this subsample. Note we use rounds 2-21 here instead of 6-21 because the analysis does not condition on the established reputation (information on the history of plays of length-5) of the players, unlike in the previous subsection} Again, we conduct the regressions separately for the \emph{slow} and \emph{fast} conditions.
Specifications (1)-(3) focus on the \emph{slow} conditions, while (4)-(6)
investigate the \emph{fast} conditions.

The key independent variable is \emph{Fraction of }$D$\emph{\ neighbors}$
_{t-1}$, which denotes the fraction of a player's neighbors whose actual
action was $D$ in the last round. In the baseline specifications (1) and
(4), we focus on \emph{Fraction of }$D$\emph{\ neighbors}$_{t-1}$, without
including any treatment dummy variables.  In specifications (2) and (5), we
include an intercept dummy variable for treatments with \emph{Reputational
uncertainty}, and interact the dummy variable with \emph{Fraction of }$D$
\emph{\ neighbors}$_{t-1}$. Finally, in specifications (3) and (6), we check
for robustness by including network metrics of the deciding player. The
remaining control variables are the same as in Tables \ref
{tab:lenient} and \ref{tab:forgiving}.

\medskip
\begin{table}[h]
	\small
	\caption{\small Random-effects logit regressions - effect of rep. uncertainty on action punishment.}
	\centering
	\begin{tabular}[c]{lccccccc}
		\hline\hline
		&\multicolumn{3}{c}{\textit{Slow}}  & \qquad  \qquad & \multicolumn{3}{c}{\textit{Fast}}  \\
		& (1) & (2) & (3) &  & (4) & (5) & (6) \\
		\hline
		\multicolumn{8}{l}{\textbf{Dependent variable: Intended defecting action} (=1 if $D$)}\\
		&&&&&&&\\
		Fraction of $D$ neighbors$_{t-1}$ & 3.872*** & 4.602*** & 4.651*** &  & 2.656*** & 3.375*** & 3.652*** \\
		& (0.397)  & (0.567)  & (0.584)  &  & (0.254)  & (0.426)  & (0.470)  \\
		Fraction of $D$ neighbors$_{t-1}$ &          & -1.478*  & -1.527** &  &          & -1.504** & -1.690** \\
		$\times$ Reputational uncertainty         &          & (0.778)  & (0.762)  &  &          & (0.645)  & (0.687)  \\
		&&&&&&&\\
		Reputational uncertainty          &          & 0.878    & 1.003    &  &          & 0.786*** & 0.813**  \\
		&          & (0.674)  & (0.649)  &  &          & (0.290)  & (0.320)  \\
		&          &          &          &  &          &          &          \\
		Degree                            &          &          & 0.253**  &  &          &          & 0.130**  \\
		&          &          & (0.113)  &  &          &          & (0.062)  \\
		Local clustering                  &          &          & -1.018   &  &          &          & 0.384    \\
		&          &          & (0.874)  &  &          &          & (0.439)  \\
		Betweenness centrality            &          &          & -8.576** &  &          &          & 1.198    \\
		&          &          & (3.417)  &  &          &          & (1.485) \\
		&&&&&&&\\
		\hline
		Control and intercept & Yes & Yes & Yes    &         & Yes & Yes & Yes              \\
		Observations          & \multicolumn{3}{c}{1985} && \multicolumn{3}{c}{2335}      \\
		\hline\hline
	\end{tabular}
	\begin{tablenotes}
		\footnotesize
		\item Notes: Standard errors (clustered at the group level) are in parentheses, {*} \(p<0.10\), {**} \(p<0.05\), {***} \(p<0.01\). Control variables include round number, age, gender, trust and risk aversion profiles, and prior exposure to game theory and economic experiments.
	\end{tablenotes}
	\label{tab:action}
\end{table}

\begin{figure}[tbp]
	\caption{\small Estimated likelihood of participants continuing to choose $D$ as
		their intended action in a round, conditioned on the fraction of defecting
		neighbors in the last round (based on (3) and (6) in Table \protect\ref{tab:action}). Vertical lines indicate 95\% confidence intervals.}
	\vspace{-20pt}
	\begin{center}
		\includegraphics[width=0.99\textwidth]{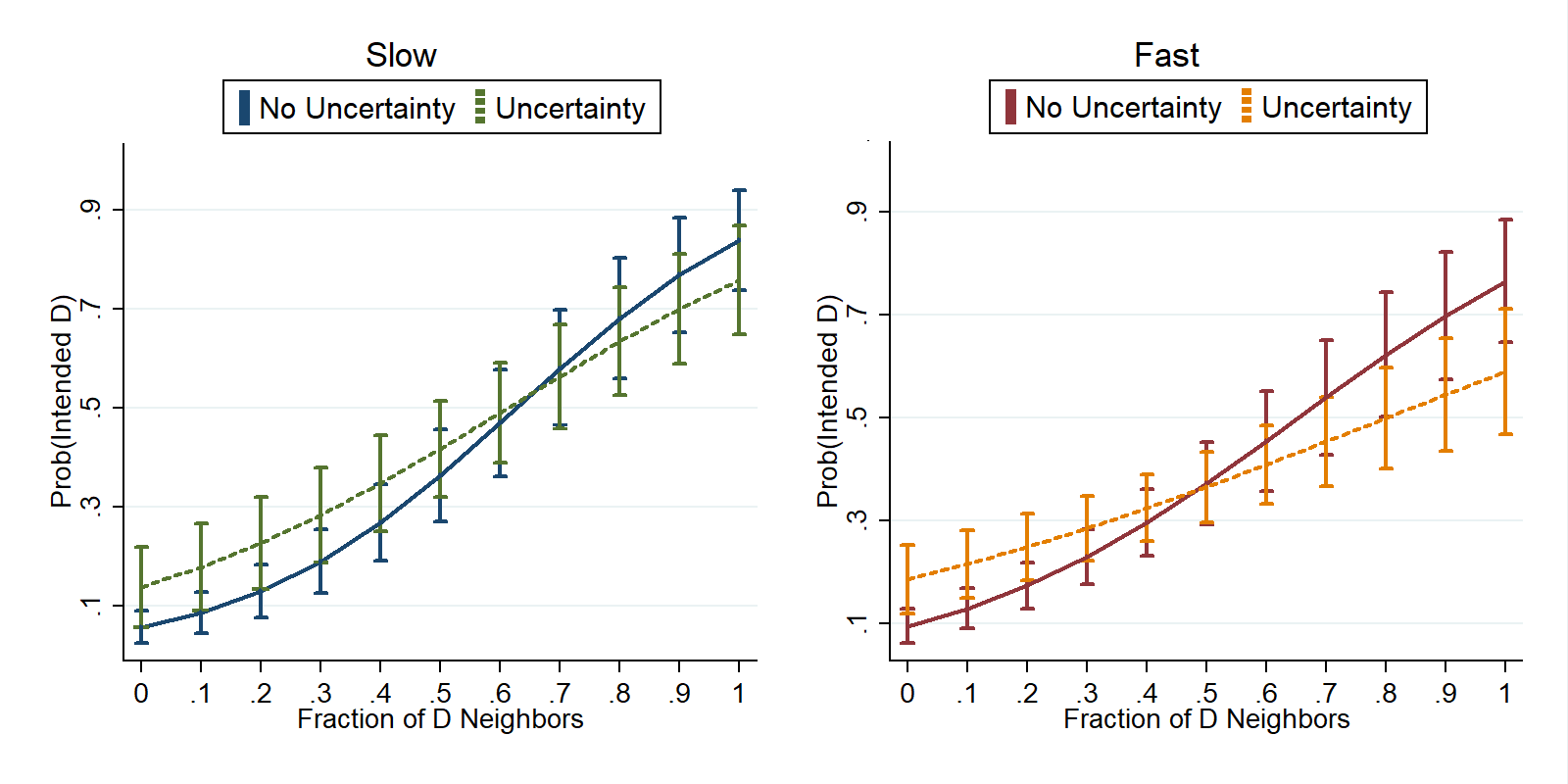}
	\end{center}
	\label{fig:action_punish}
\end{figure}

The main finding from Table \ref{tab:action} is that the probability of
choosing $D$ as the intended action in a round is significantly increasing ($
p<0.01$ for all specifications) in the fraction of defecting neighbors,
suggesting that subjects engage in action punishment. Then, specifications
(3) and (6) show that the presence of reputational uncertainty alters this
dependency, thus dampening action punishment in both \emph{slow} and \emph{fast} conditions. Specifically, with reputational uncertainty, players become less
sensitive toward past actions of their neighbors (the coefficient of the
interaction term between \emph{Reputational uncertainty }and \emph{Fraction
of }$D$\emph{\ neighbors}$_{t-1}$, $p<0.05$ in both conditions) and become more likely to defect (the coefficient of \emph{
Reputational uncertainty}, $p=0.122$ and $p<0.05$ in \emph{slow} and \emph{
fast} conditions respectively). Figure \ref{fig:action_punish} illustrates
that these effects of reputational uncertainty are consistent with the intuition discussed at the beginning of this subsection.

\bigskip

\textbf{Result 5}: \textit{In both fast and slow conditions: (i) individual
subject's probability of choosing $D$ is significantly increasing in the
fraction of defecting neighbors; (ii) reputational uncertainty makes this
probability significantly less sensitive towards the fraction of defecting
neighbors (i.e., less action punishment). }

\bigskip

\bigskip

\subsubsection{Substituting action punishment with network punishment}
At the end of Section \ref{social}, we pointed out that network punishment is more effective in a fast environment than in a slow environment. This suggests that subjects in a slow environment may primarily rely on action punishment to punish defectors. In contrast, subjects in a fast environment may substitute action punishment with the more targeted network punishment.

To examine this conjecture, we replicate the analysis in Table \ref{tab:action} but focus on the effect of the fast condition. We conduct the regression separately for treatments without reputational uncertainty and with reputational uncertainty. The independent variables of interest are the treatment dummy variable \textit{Fast} and its interaction with \textit{Fraction of $D$ neighbors$_{t-1}$}. All other specifications are the same as in Table \ref{tab:action}.

\medskip
\begin{table}[ht]
	\small
	\caption{\small Random-effects logit regressions - effect of fast condition on action punishment.}
	\centering
	\begin{tabular}[c]{lccccccc}
		\hline\hline
		&\multicolumn{3}{c}{\textit{No uncertainty}}  & \qquad  & \multicolumn{3}{c}{\textit{Reputational uncertainty}}  \\
		& (1) & (2) & (3) &  & (4) & (5) & (6) \\
		\hline
		\multicolumn{8}{l}{\textbf{Dependent variable: Intended action} (=1 if $D$)}\\
		&&&&&&&\\
		Fraction of $D$ neighbors$_{t-1}$ & 3.900*** & 4.667***  & 4.629*** && 2.363*** & 3.073*** & 3.176*** \\
		& (0.273)  & (0.311)   & (0.329)  && (0.322)  & (0.295)  & (0.262)  \\
		Fraction of $D$ neighbors$_{t-1}$ $\times$ \textit{Fast} &          & -1.262*** & -0.958*  &&          & -1.216** & -1.210** \\
		 &          & (0.487)   & (0.580)  &&          & (0.475)  & (0.497)  \\
		\textit{Fast}    &          & 0.514     & 0.531    &&          & 0.504   & 0.563   \\
		&          & (0.323)   & (0.333)  &         & & (0.298)  & (0.331)  \\
		&&&&&&&\\
		\hline
		Network metrics       & No & No & Yes    &         & No & No & Yes              \\
		Control and intercept & Yes & Yes & Yes    &         & Yes & Yes & Yes              \\
		Observations          & \multicolumn{3}{c}{1985} && \multicolumn{3}{c}{2335}      \\
		\hline\hline
	\end{tabular}
	\begin{tablenotes}
		\footnotesize
		\item Notes: Standard errors (clustered at the group level) are in parentheses, {*} \(p<0.10\), {**} \(p<0.05\), {***} \(p<0.01\). Network metrics include degree, local clustering coefficient, and centrality. Control variables include round number, age, gender, trust and risk aversion profiles, and prior exposure to game theory and experiments.
	\end{tablenotes}
	\label{tab:action dynamic}
\end{table}

The main finding from Table \ref{tab:action dynamic} comes from specifications (3) and (6). Holding the presence of reputational uncertainty constant, introducing the fast condition makes players less
sensitive towards past actions of their neighbors (the coefficient of the
interaction term between \emph{Fast }and \emph{Fraction
of }$D$\emph{\ neighbors}$_{t-1}$, $p<0.10$ without uncertainty and $p<0.05$ with uncertainty). This is consistent with the intuition that a faster-updating environment induces players to substitute away from action punishment.

\bigskip

\textbf{Result 6}: \textit{Holding constant the presence of reputational uncertainty: (i) individual
subject's probability of choosing $D$ is significantly increasing in the
fraction of defecting neighbors; (ii) fast-changing social environment makes this
probability significantly less sensitive towards the fraction of defecting
neighbors (i.e., substituting away from action punishment). }

\bigskip

\subsubsection{Reputational uncertainty promotes opportunistic play}
Apart from its effect on action punishment, reputational uncertainty could
undermine overall cooperation by promoting \textquotedblleft opportunistic
play\textquotedblright . Specifically, in the presence of reputational
uncertainty, cooperators could occasionally defect, knowing that others
would be unable to ascertain whether or not the defection is intentional,
i.e., they can hide their occasional defections \textquotedblleft behind the
veil of noise\textquotedblright. For the purpose of analysis, we define
this behavior as follows.

\begin{itemize}
\item \textbf{Opportunistic Play. }The act of a player choosing $D$
as the intended action after establishing a history of high cooperative play
(i.e., actual action was $C$ at least four times in the last five rounds).\footnote{
Note that this definition of high cooperative play is consistent with the
categorization in Section \ref{leniency and forgiveness}. Our results remain
robust if we define this based on exactly 4 $C$ instead.}
\end{itemize}

\begin{figure}[tbp]
\caption{\small Probability of intended action being $D$ given specific history of own actual actions.}
\vspace{-20pt}
\begin{center}
\includegraphics[width=0.6\textwidth]{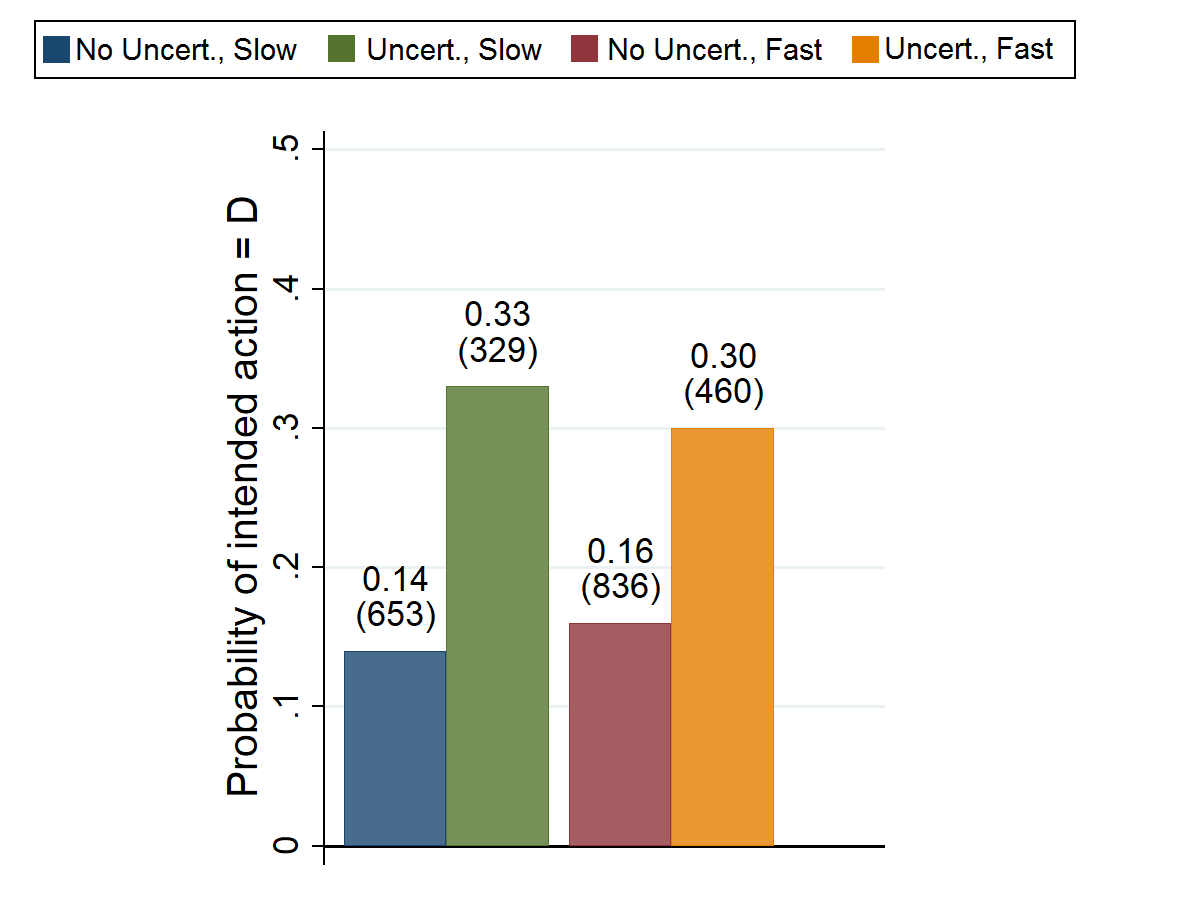}
\end{center}
\vspace{-20pt}
\caption*{\footnotesize Notes: we focus on players who had at least 4 occurrence of the actual action being $C$
out of the last 5 rounds. We report the average values over rounds 6-21. The number of
observations are in parentheses.}
\label{fig:opp}
\end{figure}

Figure \ref
{fig:opp} reports the average probability of opportunistic play in each treatment
over rounds 6-21. We then apply the Mann-Whitney rank-sum test (\emph{MW})
to evaluate the effect of reputational uncertainty, holding constant the rate of
network updating. We find that reputational uncertainty
significantly increases the likelihood of opportunistic play from 0.16 to
0.30 in the \emph{fast} condition ($p=0.021$, $MW$), and from 0.14 to 0.33
in the \emph{slow} condition ($p=0.009$, $MW$).

\bigskip

\textbf{Result 7}:\textit{\ In both fast and slow conditions,
reputational uncertainty significantly promotes opportunistic play.}

\bigskip

The rise in opportunistic play in Result 7 provides a complementary explanation for the drop in the
overall cooperation level when reputational uncertainty is present. Previous
experimental research on repeated games has found that the presence of reputational
uncertainty can result in a higher share of players adopting non-cooperative
strategies like \textquotedblleft always defecting\textquotedblright\ or a
\textquotedblleft defection-inclined version of
tit-for-tat\textquotedblright\ (\cite{fudenberg2012slow}, \cite{aoyagi2019impact}).
While our set-up differs from these studies, we have obtained a similar
result: opportunistic play, which can be interpreted as a type of
non-cooperative strategy, becomes more common when there is reputational
uncertainty.

\section{Conclusion}

We provide empirical evidence regarding cooperation in an uncertain world
with individuals receiving opportunities to update their interaction partners. Our
experiment demonstrates that the presence of noise, defined as uncertainty
on whether other people's observed actions correspond to their intended
actions, affects individual behavior along the network as well as the action
dimension. We further show that the extent of the impact of noise depends
fundamentally on the frequency at which individuals receive opportunities to
propose and/or remove links with others. In doing so, we extend our
knowledge on the characterization of behavior under noise with endogenous
network formation.

Our study raises several questions that suggest exciting
possibilities for follow-up studies as future research. First, we do not
impose an explicit cost on forming, maintaining, or removing links in our
experiment.\footnote{
Real-world relationships take time to form, and maintenance of such links
involves time and effort, both of which are costly.} Exploring the effect of
such cost might provide us with additional insights than mere opportunity
costs of not interacting as in our setup.

Second, there are few studies that
investigate the effect of allowing communication on cooperation in a two-person repeated
PD game \citep{arechar2017m} and a partnership game \citep{embrey2013experimental}. It would be fruitful
to study the effect of communication in our framework.

Third, viewing cooperation as a risky choice, the strategic decisions made by individuals affect the distribution of income and wealth in society. In a static, non-strategic, and non-networked context, \cite{cappelen2013just} demonstrate that, even though many subjects focus exclusively on ex-ante opportunities, most favor some redistribution ex-post, and several individuals make a distinction between ex-post inequalities that reflect differences in luck and ex-post inequalities that reflect differences in choices. It would be interesting to explore the fairness and inequality consequences of the action and link proposal/removal decisions in our setup.

Finally, noisy
information is publicly observable in our setting, so our game has an
imperfect public monitoring structure. However, it is possible that players
do not even know what signals other players have observed about their own
play, thereby making the monitoring structure private. \cite
{aoyagi2019impact} and \cite{kayaba2016accuracy} study cooperation in two-person
experimental games with such a monitoring structure. Our setup could be extended to
include this type of challenging monitoring technology or, more generally, alternative formulations of reputational uncertainties.

{\small
\bibliographystyle{chicago}
\bibliography{references}
}

\clearpage

\appendix\pagenumbering{arabic}
\setcounter{table}{0}
\begin{spacing}{1.2}
\renewcommand{\thetable}{A\arabic{table}}

\begin{center}
	{\LARGE Appendix: Cooperation and Punishment Mechanisms in Uncertain and Dynamic Social Networks}
\end{center}

\section{Demographics} \label{sec: demographics}

Table A1 summarizes the main socio-demographic characteristics of our participants. We verified that there are no significant differences ($p>0.10$) across the four treatments in terms of these socio-demographic characteristics.

\begin{table}[ht]
	\caption{Average values of demographic variables.}
	\scalebox{0.9}{
		\begin{threeparttable}
			\renewcommand{\TPTminimum}{\linewidth}
			\makebox[\linewidth]{
				\begin{tabular}{lllll}
					\hline\hline
					Treatment & \textit{(No Uncertainty,} & \textit{(Uncertainty,} & \textit{(No Uncertainty,} & \textit{(Uncertainty,}\\
					& \textit{Slow)} & \textit{Slow)} & \textit{Fast)} & \textit{Fast)}\\\hline
					Age & 20.91 & 21.31 & 20.84 & 21.01\\
					Gender (Female) & 57\% & 45\% & 59\% & 49\%\\
					Risk Aversion & 5.42 & 5.62 & 5.91 & 6.03\\
					Game Theory & 25\% & 24\% & 23\% & 22\%\\
					Prior Participation & 63\% & 46\% & 46\% & 47\%\\
					Trust & 25\% & 17\% & 28\% & 20\%\\\hline
			\end{tabular}}
		\label{tab:appx demographics}
	\end{threeparttable}}
	\begin{tablenotes}
		\footnotesize
		\item \emph{Risk Aversion} shows the score from Holt and Laury (2002) risk-elicitation test, in which a score of 0 indicates minimal risk aversion; \emph{Trust} shows the percentage of participants indicating ``yes'' in the standard interpersonal trust question taken from the World Values Survey (WVS).
	\end{tablenotes}
\end{table}

\section{Additional results}
\subsection{Alternative definitions for leniency and forgiveness}
\label{sec: alt definition}
Our results remain robust if we redefine \textit{Leniency} and \textit{Forgiveness} in Section \ref{social} by specifying ``history of low cooperative play'' as actual action (of the target neighbor/potential neighbor) being $C$ at most two times in the last 5 rounds.

Tables \ref{tab:appx alt leniency} and \ref{tab:appx alt forgiveness} below replicate Results 2 and 3. Given the new definitions, in what follows, we focus on observations where the target neighbor/potential neighbor has actual action being $C$ at most two times in the last five rounds. All other specifications and controls used are the same as Tables \ref{tab:lenient} and \ref{tab:forgiving}. We find that the coefficient of \textit{Reputational uncertainty} in the \textit{Fast} condition remains significant ($p<0.05$) and that the results remain robust if we exclude the demographic variables and the network metrics from the regressions.

\begin{table}[ht]
	\small
	\caption{\small Random-effects logit regressions - effect of rep. uncertainty on leniency (new definition).}%
	\centering
	\begin{tabular}[c]{lccccccc}
		\hline\hline
		&\multicolumn{3}{c}{\textit{Slow}}  & \qquad  & \multicolumn{3}{c}{\textit{Fast}}  \\
		\multicolumn{8}{l}{History of the target neighbor: \qquad \qquad \qquad Actual action being $C$ at most twice}  \\
		\hline
		\multicolumn{8}{l}{\textbf{Dependent variable: Link removal} (conditional on receiving an opportunity)}\\
		&&&&&&&\\
			Reputational uncertainty & -0.330   & -0.247   & -0.240   &  & -0.662** & -0.733*** & -0.720** \\
			& (0.325)  & (0.295)  & (0.304)  &  & (0.261)  & (0.257)   & (0.286)  \\
			Own C Action$_{t-1}$       & 0.957*** & 0.913*** & 0.904*** &  & 0.393**  & 0.376**   & 0.369**  \\
			& (0.199)  & (0.204)  & (0.197)  &  & (0.183)  & (0.179)   & (0.186)  \\
			Degree                   &          &          & 0.035  &  &          &           & 0.005    \\
			&          &          & (0.117)                      &  &          &           & (0.043)  \\
			Local clustering         &          &          & 0.384   &  &          &           & -0.784** \\
			&          &          & (0.686)                      &  &          &           & (0.326)  \\
			Betweenness centrality   &          &          & -1.628 &  &          &           & -3.113** \\
			&          &          & (4.367)                      &  &          &           & (1.370) \\
		&&&&&&&\\
		\hline
		Control & No & Yes & Yes    &         & No & Yes & Yes              \\
		Intercept & Yes & Yes & Yes    &         & Yes & Yes & Yes              \\
		Observations          & \multicolumn{3}{c}{899} && \multicolumn{3}{c}{2606}      \\
		\hline\hline
	\end{tabular}
	\begin{tablenotes}
		\footnotesize
		\item Notes: Standard errors (clustered at the group level) are in parentheses, {*} \(p<0.10\), {**} \(p<0.05\), {***} \(p<0.01\). Control variables include round number, age, gender, trust and risk aversion profiles, and prior exposure to game theory and economic experiments.
	\end{tablenotes}
	\label{tab:appx alt leniency}
\end{table}

\begin{table}[ht]
	\small
	\caption{\small Random-effects logit regressions - effect of rep. uncertainty on forgiveness (new definition).}%
	\centering
	\begin{tabular}[c]{lccccccc}
		\hline\hline
		&\multicolumn{3}{c}{\textit{Slow}}  & \qquad  & \multicolumn{3}{c}{\textit{Fast}}  \\
		\multicolumn{8}{l}{History of the target neighbor: \qquad \qquad \qquad Actual action being $C$ at most twice}  \\
		\hline
		\multicolumn{8}{l}{\textbf{Dependent variable: Link proposal} (conditional on receiving an opportunity)}\\
		&&&&&&&\\
		Reputational uncertainty & 0.364                & 0.363    & 0.298     &  & 0.752**   & 0.865***  & 0.966***  \\
		& (0.329)              & (0.337)  & (0.418)   &  & (0.297)   & (0.294)   & (0.274)   \\
		Own C Action$_{t-1}$       & -0.806**             & -0.790** & -0.668**  &  & -0.418*** & -0.374*** & -0.375*** \\
		& (0.373)              & (0.361)  & (0.335)   &  & (0.111)   & (0.130)   & (0.123)   \\
		&&&&&&&\\
		Degree                   &  &          & -0.313*** &  &           &           & -0.188*** \\
		& \multicolumn{1}{c}{} &          & (0.112)   &  &           &           & (0.043)   \\
		Local clustering         &                      &          & -0.712    &  &           &           & -0.438*   \\
		&                      &          & (0.473)   &  &           &           & (0.236)   \\
		Betweenness centrality   &                      &          & 5.669     &  &           &           & 3.079*    \\
		&                      &          & (6.886)   &  &           &           & (1.625)  \\
		&&&&&&& \\
		\hline
Control & No & Yes & Yes    &         & No & Yes & Yes              \\
Intercept & Yes & Yes & Yes    &         & Yes & Yes & Yes              \\
		Observations          & \multicolumn{3}{c}{628} && \multicolumn{3}{c}{3340}      \\
		\hline\hline
	\end{tabular}
	\begin{tablenotes}
		\footnotesize
		\item Notes: Standard errors (clustered at the group level) are in parentheses, {*} \(p<0.10\), {**} \(p<0.05\), {***} \(p<0.01\). Control variables include round number, age, gender, trust and risk aversion profiles, and prior exposure to game theory and economic experiments.
	\end{tablenotes}
	\label{tab:appx alt forgiveness}
\end{table}

\clearpage
\subsection{Forgiveness in both proposal and acceptance}
\label{sec: accept}
In Section \ref{social}, our definition of \emph{Forgiveness} is based on link proposal decisions.  We can alternatively broaden the definition by considering both link proposal and link acceptance decisions. That is, strategies are forgiving if, conditional on receiving an opportunity to propose a link, the deciding player \textit{proposes a link to} or \textit{accepts a link proposal (after initially deciding not to propose)} from someone with a history of low cooperative play (i.e., whose actual action was $C$ at most once in the last five rounds).

Table \ref{tab:appx forgive accept} is analogous to Table \ref{tab:forgiving}, except that we now additionally include observations that are link acceptance decisions. In other words, we treat observations of link proposal decisions and link acceptance decisions (whenever it arises) pertaining to the potential neighbor as two separate observations. The dependent variable is the link forming decision ($=1$ if the participant sends a proposal to or accepts a link proposal from the potential neighbor). We also included a dummy variable \emph{Is acceptance decision} (=1 if the observation is a link acceptance decision).

\begin{table}[h]
	\small
	\caption{\small Random-effects logit regressions - effect of uncertainty on forgiveness (incl. acceptance decisions).}%
	\centering
	\begin{tabular}[c]{lccccc}
		\hline\hline
		&\multicolumn{2}{c}{\textit{Slow}}    &   \qquad\qquad\qquad   & \multicolumn{2}{c}{\textit{Fast}} \\
		History of the target neighbor: \qquad & {Low $C$}   & {Medium $C$}  & &  {Low $C$}  & {Medium $C$}      \\
		\hline		
		\multicolumn{6}{l}{\textbf{Dependent variable: Link proposal or acceptance} (conditional on receiving an opportunity)}\\
		&&&&&\\
		Reputational uncertainty & 1.784     & 0.404     &  & 1.376***  & 0.815***  \\
		& (3.276)   & (0.399)   &  & (0.510)   & (0.313)   \\
		Own C Action$_{t-1}$       & -1.222    & -1.311*** &  & -0.064    & -0.227    \\
		& (0.823)   & (0.339)   &  & (0.159)   & (0.169)   \\
		Is acceptance decision   & -1.615*** & -1.845*** &  & -0.937*** & -1.219*** \\
		& (0.453)   & (0.259)   &  & (0.144)   & (0.145)  \\
		&&&&&\\ \hline
		Network metrics & Yes & Yes    &         & Yes & Yes              \\
		Control and intercept & Yes & Yes    &         & Yes & Yes              \\
		Observations          & 591    &  569   &  & 3227 & 3545           \\
		\hline\hline
	\end{tabular}
	\begin{tablenotes}
		\footnotesize
		\item Notes: \textit{Low $C$}- $C$ was the actual action at most one time in last 5 rounds; \textit{Medium $C$}- $C$ was the actual action two or three times. Standard errors (clustered at the group level) are in parentheses, {*} \(p<0.10\), {**} \(p<0.05\), {***} \(p<0.01\). Network metrics include degree, local clustering coefficient, and betweenness centrality. Control variables include round number, age, gender, trust and risk aversion profiles, and prior exposure to game theory and economic experiments.
	\end{tablenotes}
	\label{tab:appx forgive accept}%
\end{table}

There are two main findings from Table \ref{tab:appx forgive accept}. First, Result 3 remains robust as indicated by the coefficient of \textit{Reputational uncertainty} in the fast condition ($p<0.01$ for both \textit{Low $C$} and \textit{Medium $C$}, respectively). Second, all else being equal, players are more likely to say ``yes''  when faced with a decision to send a link proposal initially than to say ``yes'' in an acceptance decision (after initially deciding not to propose), as indicated by the coefficient of \textit{Is acceptance decision} ($p<0.01$).

To illustrate the second finding, Figure A1 is analogous to Figure \ref{fig:forgiving} but focuses on acceptance decisions. Notice that the number of observations is low compared to the decisions of sending a proposal in Figure \ref{fig:forgiving}. Also, the likelihood of players saying ``yes'' to an incoming proposal (after not sending a proposal initially) is generally below 0.50 across all play histories of the potential neighbor.

\begin{figure}[H]
	\caption*{\small Figure A1: Average likelihood to accept an incoming proposal (after not proposing initially).}
	\vspace{-20pt}
	\begin{center}
		\includegraphics[width=0.95\textwidth]{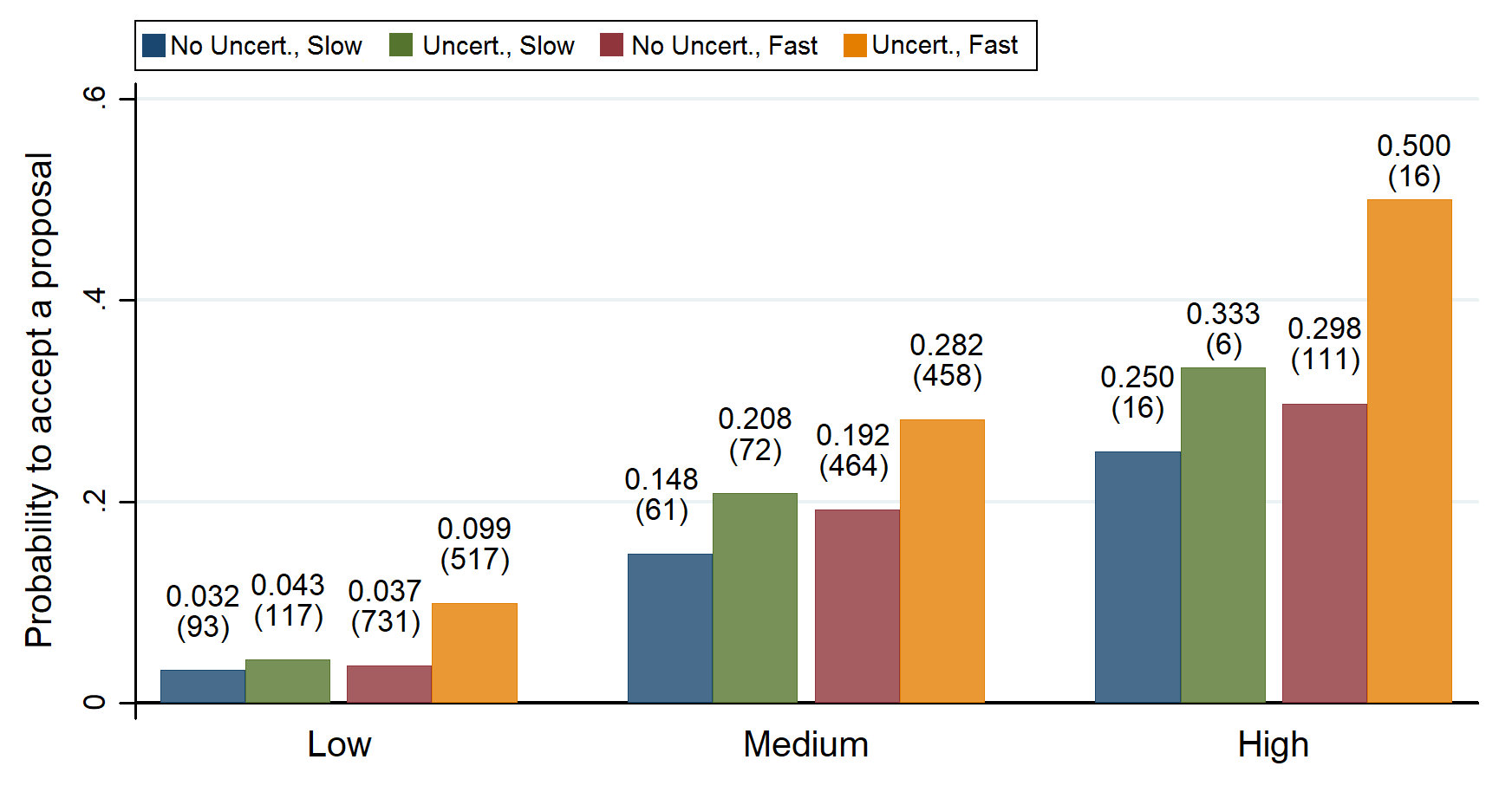}
	\end{center}
	\vspace{-20pt}
	\caption*{\footnotesize Notes: Separated by the last five rounds' history, where \textit{Low}: the actual action was $C$ at most once in the last five rounds;
		\textit{Medium}: $C$ was the actual action two or three times; \textit{High}%
		: $C$ was the actual action more than three times. We report the average values over rounds 6-21, with the number of
		observations in parentheses.}
	\label{fig:accept}
\end{figure}

\subsection{Effect of network updating rate on leniency and forgiveness}
\label{sec: dynamic effect}
Tables \ref{tab:appx dynamic leniency} and \ref{tab:appx dynamic forgiveness} are similar to the analysis in Section \ref{leniency and forgiveness}, except that here we focus on the effect of \textit{Fast} condition. All other specifications and controls used are the same as Tables \ref{tab:lenient} and \ref{tab:forgiving}. Introducing \textit{fast} condition significantly raises leniency ($p<0.05$), regardless of the level of reputational uncertainty. Meanwhile, it has no significant impact on forgiveness ($p>0.1$).

\begin{table}[ht]
	\small
	\caption{\small Random-effects logit regressions - effect of network updating rate on leniency.}%
	\centering
	\begin{tabular}[c]{lccccc}
		\hline\hline
		&\multicolumn{2}{c}{\textit{No uncertainty}}  & \qquad\qquad\qquad  & \multicolumn{2}{c}{\textit{Reputational uncertainty}}  \\
		History of the target neighbor: \qquad & {Low $C$}   & {Medium $C$}  & &  {Low $C$}  & {Medium $C$}      \\
		\hline		
		\multicolumn{6}{l}{\textbf{Dependent variable: Link removal} (conditional on receiving an opportunity)}\\
		&&&&&\\
		\textit{Fast} & -1.032*** & 0.023   &  & -1.377*** & -0.604**  \\
		& (0.319)   & (1.822) &  & (0.328)   & (0.294)   \\
		Own C Action$_{t-1}$       & 0.694*    & 0.494   &  & 0.583***  & 0.421     \\
		& (0.421)   & (0.348) &  & (0.175)   & (0.262)   \\
		&&&&&\\
		Degree                   & -0.009    & 0.024   &  & 0.071     & -0.142*** \\
		& (0.083)   & (0.319) &  & (0.048)   & (0.047)   \\
		Local clustering         & 1.791**   & 0.422   &  & -1.374**  & 0.628     \\
		& (0.805)   & (1.601) &  & (0.603)   & (0.602)   \\
		Betweenness centrality   & -0.806    & -0.148  &  & -6.866*** & -2.999    \\
		& (2.474)   & (4.996) &  & (2.440)   & (2.577)  \\
		&          &          &  &          &         \\ 	\hline
		Control and intercept & Yes & Yes    &         & Yes & Yes              \\
		Observations          & 565    &  1490   &  & 952 & 2999           \\
		\hline\hline
	\end{tabular}
	\begin{tablenotes}
		\footnotesize
		\item Notes: \textit{Low $C$}- $C$ was the actual action at most one time in last 5 rounds; \textit{Medium $C$}- $C$ was the actual action two or three times. Standard errors (clustered at the group level) are in parentheses, {*} \(p<0.10\), {**} \(p<0.05\), {***} \(p<0.01\). Control variables include round number, age, gender, trust and risk aversion profiles, and prior exposure to game theory and economic experiments.
	\end{tablenotes}
\label{tab:appx dynamic leniency}
\end{table}

\begin{table}[ht]
	\small
	\caption{\small Random-effects logit regressions - effect of network updating rate on forgiveness.}%
	\centering
	\begin{tabular}[c]{lccccc}
		\hline\hline
		&\multicolumn{2}{c}{\textit{No uncertainty}}  & \qquad\qquad\qquad  & \multicolumn{2}{c}{\textit{Reputational uncertainty}}  \\
		History of the potential neighbor: \qquad & {Low $C$}  & {Medium $C$}  & &  {Low $C$}    & {Medium $C$} \\
		\hline
		\multicolumn{6}{l}{\textbf{Dependent variable: Link proposal} (conditional on receiving an opportunity)}\\
		&&&&&\\
		\textit{Fast} & 0.130     & -0.526    &  & -0.330   & -0.054    \\
		& (0.477)   & (0.393)   &  & (0.547)  & (0.251)   \\
		Own C Action$_{t-1}$       & -0.595**  & -0.736**  &  & -0.296   & -0.357*** \\
		& (0.233)   & (0.339)   &  & (0.213)  & (0.137)   \\
		&&&&&\\
		Degree                   & -0.161    & -0.211*** &  & -0.221** & -0.126**  \\
		& (0.101)   & (0.047)   &  & (0.101)  & (0.060)   \\
		Local clustering         & -1.857*** & -0.541    &  & -0.405   & -0.198    \\
		& (0.599)   & (0.356)   &  & (0.315)  & (0.364)   \\
		Betweenness centrality   & 0.190     & 5.122*    &  & 7.900**  & 5.580*    \\
		& (3.020)   & (2.645)   &  & (3.713)  & (2.888)  \\
		&&&&&\\
		Control and intercept & Yes & Yes    &         & Yes & Yes              \\
		Observations          & 1211    &  1228   &  & 1149 & 1831           \\
		\hline\hline
	\end{tabular}
	\begin{tablenotes}
		\footnotesize
		\item Notes: \textit{Low $C$}- $C$ was the actual action at most one time in last 5 rounds; \textit{Medium $C$}- $C$ was the actual action two or three times. Standard errors (clustered at the group level) are in parentheses, {*} \(p<0.10\), {**} \(p<0.05\), {***} \(p<0.01\). Control variables include round number, age, gender, trust and risk aversion profiles, and prior exposure to game theory and economic experiments.
	\end{tablenotes}
\label{tab:appx dynamic forgiveness}
\end{table}

\clearpage
\subsection{Leniency and type of link formation}
\label{sec: leniency and linknature}
In our experimental environment, broadly speaking, there are three ways through which a link between two players can be formed: (i) imposed by default starting from round 1; (ii) formed through mutual proposal; (iii) formed through a mixture of proposal and acceptance decisions of the two players. We summarize the properties of each type of link below (those that lasted for at least 1 round):

\begin{table}[h]
	\small
	\centering
	\begin{tabular}{llll}
		\hline\hline
		& Default links\vspace{2pt}  \qquad\qquad & Mutually-proposed links \qquad & Proposal-acceptance links \qquad \\ \hline
		Link duration \qquad  \vspace{2pt} 		& 11.56 (8.23) & 6.75 (5.40) & 5.64 (4.76) \\
		Observations \qquad \vspace{1pt}           & 1972 & 1070 &  524 \\
		\hline\hline
	\end{tabular}%
	\begin{tablenotes}
		\footnotesize
		\item Note: we report mean values in all treatments throughout all rounds with standard errors in parentheses.
	\end{tablenotes}
\end{table}

Observe that the second type of links tends to last longer than the third type. This suggests that players making link removal decisions may treat links formed from their own proposal differently than links they accepted but did not initially propose. To explore this question, Table \ref{tab:lenient linknature} below replicates the leniency analysis in Table \ref{tab:lenient} but includes the following additional dummy variables: (i) \textit{Was a default link} (=1 if the link was imposed from round 1), and (ii) \textit{Was an own accepted link} (=1 if the link was formed from the deciding player's acceptance after not proposing a link initially). This specification means that the baseline case (when both dummy variables are equal to zero) refers to links formed from the deciding player's proposal.

\begin{table}[h]
	\small
	\caption{\small Random-effects logit regressions - effects of rep. uncertainty and types of links on leniency.}%
	\centering
	\begin{tabular}[c]{lccccc}
		\hline\hline
		&\multicolumn{2}{c}{\textit{Slow}}    &   \qquad\qquad\qquad   & \multicolumn{2}{c}{\textit{Fast}} \\
		History of the target neighbor: \qquad \qquad & {Low $C$}   & {Medium $C$}  & &  {Low $C$}  & {Medium $C$}      \\
		\hline		
		\multicolumn{6}{l}{\textbf{Dependent variable: Link removal} (conditional on receiving an opportunity)}\\
		&&&&&\\
				Reputational uncertainty & -0.404   & -0.374   &  & -0.907** & -0.784*** \\
				& (0.416)  & (0.348)  &  & (0.373)  & (0.191)   \\
				Own C Action$_{t-1}$       & 1.449*** & 0.737*** &  & 0.347*   & 0.403     \\
				& (0.398)  & (0.156)  &  & (0.184)  & (0.255)   \\
				Was an own accepted link        & 1.304    & 1.092    &  & 0.373*   & 0.427***  \\
				& (1.964)  & (1.234)  &  & (0.220)  & (0.146)   \\
				Was a default link         & 1.279**  & 0.712    &  & -0.059   & -0.541*** \\
				& (0.553)  & (0.562)  &  & (0.101)  & (0.124)  \\
		&&&&&\\
		Network metrics       & Yes & Yes    &         & Yes & Yes              \\
		Control and intercept & Yes & Yes    &         & Yes & Yes              \\
		Observations          & 508    &  754   &  & 1009 & 3735           \\
		\hline\hline
	\end{tabular}
	\begin{tablenotes}
		\footnotesize
		\item Notes: \textit{Low $C$}- $C$ was the actual action at most one time in last 5 rounds; \textit{Medium $C$}- $C$ was the actual action two or three times. Standard errors (clustered at the group level) in parentheses, {*} \(p<0.10\), {**} \(p<0.05\), {***} \(p<0.01\). Network metrics include degree, local clustering coefficient, and centrality. Control variables include round number, age, gender, trust and risk aversion profiles, and prior exposure to game theory and experiments.
	\end{tablenotes}
	\label{tab:lenient linknature}%
\end{table}

There are two main findings from Table \ref{tab:lenient linknature}. First, Result 2 remains robust as indicated by the coefficient of \textit{Reputational uncertainty} in the fast condition ($p<0.05$ and $p<0.01$ for \textit{Low $C$} and \textit{Medium $C$}, respectively). Second, all else being equal, in \textit{fast} condition, players are more likely to remove links they accepted but did not initially propose (relative to those that they proposed), as indicated by the coefficient of \textit{Was an own accepted link} ($p<0.1$ and $p<0.01$ for \textit{Low $C$} and \textit{Medium $C$}, respectively).

\subsection{Classification of Type-C and Type-D players}
\label{sec: type}
In Section \ref{network}, we classified players as either type-C or type-D according to their intended action in the first round. Table \ref{tab:appx type tabulate} summarizes the key properties of type-C and type-D players in each treatment. To check the statistical significance of the observed differences between these two types of players, we adopt the Wilcoxon paired sign-rank test for each treatment, with eight observations per treatment.

For each treatment, we find that the difference is statistically significant ($p<0.05$) for \textit{Intended action}, \textit{Average degree}, \textit{IH}, \textit{Homophily index}, and \textit{Centrality}. The only exceptions are the two treatments with reputational uncertainties, whereby the difference between type-C and type-D players in homophily index and centrality are statistically insignificant ($p>0.1$).

\begin{table}[ht]
	\small
	\caption{\small Descriptions of Type-C and Type-D players.}%
	\centering
	\begin{tabular}[c]{l|c|cccc}
		\hline\hline
			& & \multicolumn{2}{c}{Slow}  & \multicolumn{2}{c}{Fast}        \\
			& & No uncertainty & Uncertainty & No uncertainty & Uncertainty \\ \hline
		    &&&&& \\
			Number of players         & C & 6.625  & 6.750  & 6.875  & 6.750  \\
			& D & 5.375  & 5.250  & 5.125  & 5.250  \\
			&&&&& \\
			Intended action (=1 if C) & C & 0.691  & 0.538  & 0.770  & 0.666  \\
			& D & 0.482  & 0.304  & 0.559  & 0.424  \\
			&&&&& \\
			Average degree            & C & 7.872  & 7.639  & 6.821  & 7.000  \\
			& D & 7.021  & 6.782  & 5.324  & 6.327  \\
			&&&&& \\
			IH                        & C & 0.061  & -0.024 & 0.122  & -0.006 \\
			& D & -0.059 & -0.153 & -0.132 & -0.116 \\
			&&&&& \\
			Homophily index           & C & 0.581  & 0.541  & 0.625  & 0.556  \\
			& D & 0.415  & 0.359  & 0.352  & 0.373  \\
			&&&&& \\
			Centrality                & C & 0.038  & 0.038  & 0.051  & 0.041  \\
			& D & 0.028  & 0.031  & 0.035  & 0.039  \\
		\hline\hline
	\end{tabular}
	\begin{tablenotes}
		\footnotesize
		\item Note: we report mean values of these variables over rounds 2-21.
	\end{tablenotes}
	\label{tab:appx type tabulate}
\end{table}

\clearpage

\begin{small}

\section{Experimental Instructions for \textit{(Uncertainty, Fast)} treatment} \label{sec: instructions}
\setlength{\parskip}{6pt}
\setlength\parindent{0pt}

Welcome to all of you! You are now taking part in an interactive study on decision making. Please pay attention to the information provided here and make your decisions carefully. If at any time you have questions to ask, please \underline{raise your hand} and we will \underline{attend to you in private}.

Please note that {\bf unauthorized communication is prohibited}. Failure to adhere to this rule would force us to stop this study and you may be held liable for the cost incurred in this experiment.

Your participation in this study is voluntary. You will receive SGD {\bf 2} show-up fee for participating in this study. You may decide to leave the study at any time. Unfortunately, if you withdraw before you complete the study, we can only pay you for the decisions that you have made up to the time of withdrawal, which could be substantially less than you will earn if you complete the entire study.

The amount of your earnings from this study depends on the decisions you and others make and also on chance. At the end of this session, your earnings will be paid to you privately and in cash. They will be contained in an envelope (indicated with your unique user ID).

\noindent\rule{\textwidth}{0.4pt}
\begin{center}
	\bf General Information
\end{center}
Each of you will be given a unique user ID and it will be clearly stated on your computer screen. At the end of the study, you will be asked to fill in your user ID and other information pertaining to your earnings from this study in the payment receipt. {\bf Please fill in the correct user ID to make sure that you will get the correct amount of payment}.

Rest assured that your {\bf anonymity will be preserved} throughout the study. You will never be aware of the personal identities of other participants during or after the study. Similarly, other participants will also never be aware of your personal identities during or after the study. You will only be identified by your user ID in our data collection. All information collected will {\bf strictly be kept confidential} for the sole purpose of this study.

\begin{center}
	\bf Specific Information
\end{center}
The total duration of this study is approximately 2 hours.

Your total earnings	= earnings from Part I + earnings from Part II + show up fee (S$\$ 2 $)

All incentives will be denominated in {\bf experimental dollars} (expressed as {\bf ECU}). The exact conversion rate will be detailed later.

\noindent\rule{\textwidth}{0.4pt}
\begin{center}
	\bf Part I
\end{center}

You and 11 other participants will engage in several rounds of interactions. Each round is the same and consists of 3 Stages.

In Stages 1 and 2, you can form and delete links with any of the other participants, which we will explain shortly. If you are linked with another participant then this participant is a ``neighbor''.

In Stage 3, you choose either action A or action B, and the choice of action A or B applies to all your neighbors. Action A is color-coded in green and action B is color-coded in blue. The colors are only a visual aid to distinguish the actions and have no meaning. You get points for the action you choose and the action each of the other participants chooses, in the following way:

You get 0 points if you are not linked with another participant regardless of your choice of action.

The number of points you get depends on the actions you and your neighbor choose, according to the table below:

\begin{center}
	\begin{center}
		\begin{tabular}{|c|c|c|c|}
			\hline
			\multicolumn{4}{|c|}{Neighbor} \\ \hline
			\multirow{3}{4em}{You} &                            & \textcolor[rgb]{0.00,0.50,0.25}{A}    & \textcolor[rgb]{0.00,0.00,1.00}{B} \\ \cline{2-4}
			& \textcolor[rgb]{0.00,0.50,0.25}{A} & 3                             & -5 \\  \cline{2-4}
			& \textcolor[rgb]{0.00,0.00,1.00}{B}        & 5                             & -3 \\ \hline
		\end{tabular}
	\end{center}
\end{center}

This is what the table says:
\begin{itemize}
	\setlength\itemsep{0.0001em}
	\item If you choose \textcolor[rgb]{0.00,0.50,0.25}{A} and the neighbor chooses \textcolor[rgb]{0.00,0.50,0.25}{A}, you get 3 points
	\item If you choose \textcolor[rgb]{0.00,0.50,0.25}{A} and the neighbor chooses \textcolor[rgb]{0.00,0.00,1.00}{B}, you get -5 points
	\item If you choose \textcolor[rgb]{0.00,0.00,1.00}{B} and the neighbor chooses \textcolor[rgb]{0.00,0.50,0.25}{A}, you get 5 points
	\item If you choose \textcolor[rgb]{0.00,0.00,1.00}{B} and the neighbor chooses \textcolor[rgb]{0.00,0.00,1.00}{B}, you get -3 points
\end{itemize}

At the end of each round you will see a summary of the number of points you get with each of the other participants.

You will play 25 rounds of this game for sure. After round 25, there is a 50$\%$ chance that the game will terminate in the following round. In other words, in every round after round 25 the experimenter will ``flip a coin'' and if the outcome is ``Heads'' then the game will terminate.

\begin{center}
	\bf Stage 1 - Link Decisions
\end{center}
In Stage 1, you will have the opportunity to (i) form a link with another player who is not linked to you or (ii) to delete an existing link with one of your partner(s). However, you will not have the complete freedom to form or to delete links. In particular, 33 random pairs of players (out of the 66 possible pairs of players in this network) will be selected to be updated such that:

(i) \quad If you and another player that is selected are not linked currently, you will need to decide whether to propose a new link between you two.

(ii) \quad If you and another player that is selected are linked currently, you will need to decide if you would like to delete the link.

For illustration purposes, the screenshots from Round 4 are shown (see also the attachment (stage 1).		

\underline{Top-left of the screen}

You will see the following figure that visualizes your current neighbors.

\begin{figure}[H]
	\centering
	\includegraphics[scale=0.8]{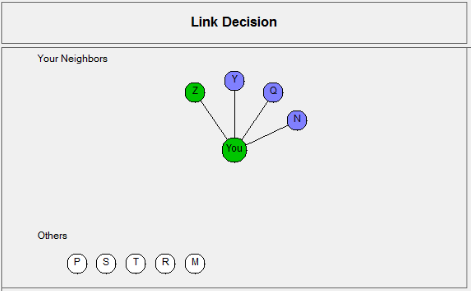}
\end{figure}

The neighbors you have at the start of each round are the same as the neighbors you had at the end of the previous round. A link means that if you are linked to participant Z then participant Z is linked to you.

Each circle denotes a participant with an ID. The ``You'' circle is always positioned at the center. For example, above you are linked with Z, Y, Q, and N. The color of the circles represents the last actual action (\textcolor[rgb]{0.00,0.50,0.25}{A} or \textcolor[rgb]{0.00,0.00,1.00}{B}) that has been chosen by each participant.

Note that in Round 1 all participants are connected to each other.

\underline{Center of the screen}

This table allows you to make your decisions to {\bf delete} links in Stage 1. The first row of the table (``Your Last 5 Actions'') reminds you of the actions you have chosen in the last 5 rounds. For example, the sequence below means that you chose action \textcolor[rgb]{0.00,0.50,0.25}{A} in the last round (the {\bf leftmost}), action \textcolor[rgb]{0.00,0.50,0.25}{A} two rounds ago (the {\bf second-leftmost} slot), action \textcolor[rgb]{0.00,0.50,0.25}{A} three rounds ago (the {\bf middle}), and action \textcolor[rgb]{0.00,0.00,1.00}{B} four rounds ago (the {\bf second rightmost}), and action \textcolor[rgb]{0.00,0.00,1.00}{B} five rounds ago (the {\bf rightmost}).

\begin{figure}[H]
	\centering
	\includegraphics[scale=0.8]{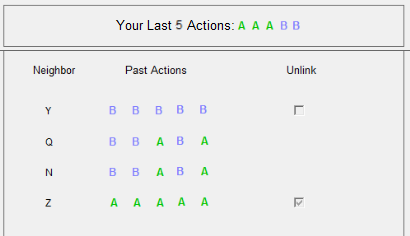}
\end{figure}

For the pairs of players selected to be updated, a box will be shown next to the ID of the participant. You can delete a link to the neighbor by ticking the corresponding box under the ``Unlink'' column. There is no limit to how many boxes among the shown boxes you can tick, and you can also choose not to delete any link by leaving the shown boxes unticked.

The table has 3 columns, from left to right:
\begin{itemize}
	\setlength\itemsep{0.0001em}
	\item {\bf ``Neighbor''}: lists the IDs of the participants who are currently linked with you
	\item {\bf ``Past actions''}: lists the last 5 actual actions by each of your neighbors
	\item {\bf ``Unlink''}: allows you to delete your link to the neighbors that have been randomly selected by the system. For example, ticking the box in the last row will delete your link to participant Z
\end{itemize}

The color-codings of the other participants' actions are the same as before. The green letter \textcolor[rgb]{0.00,0.50,0.25}{A} denotes the choice of action \textcolor[rgb]{0.00,0.50,0.25}{A}. The blue letter \textcolor[rgb]{0.00,0.00,1.00}{B} denotes the choice of action \textcolor[rgb]{0.00,0.00,1.00}{B}. The ``-'' denotes that the participant did not have any neighbor in that round, and therefore the participant did not choose an action.

\underline{Right of the screen}

This table allows you to make your decisions to {\bf form} links in Stage 1.

\begin{figure}[H]
	\centering
	\includegraphics[scale=0.8]{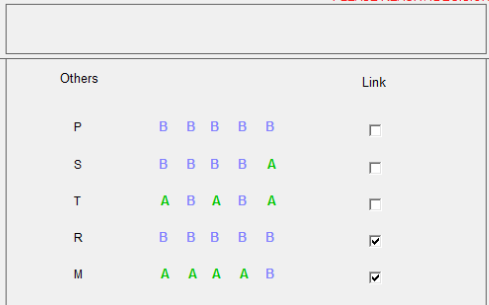}
\end{figure}

For the pairs of players selected to be updated, a box will be shown next to the ID of the participant. You can propose a link to the participant by ticking the corresponding box under the ``Link'' column. There is no limit to how many boxes among the shown boxes you can tick, and you can also choose not to propose any link by leaving all the boxes unticked.

The table has 3 columns, from left to right:
\begin{itemize}
	\setlength\itemsep{0.0001em}
	\item {\bf ``Others''}: lists the IDs of the participants who are not linked with you
	\item {\bf ``Past actions''}: lists the last 5 actual actions by each of these participants
	\item {\bf ``Link''}: allows you to propose a link to the participants randomly selected by the system you are not linked with. For example, ticking the box in the last row will propose a link to participant M
\end{itemize}

\underline{Bottom-right of the screen}

Note that the participants whom you can propose or delete links with are decided randomly by the computer. The text reminds you about the decisions you need to make and lists the participants whom you can propose or delete links with. You can propose new links and/or delete existing links, or you can choose not to propose and/or delete any links. In either case, you need to click the ``Submit'' button to confirm your decisions.

\begin{figure}[H]
	\centering
	\includegraphics[scale=0.7]{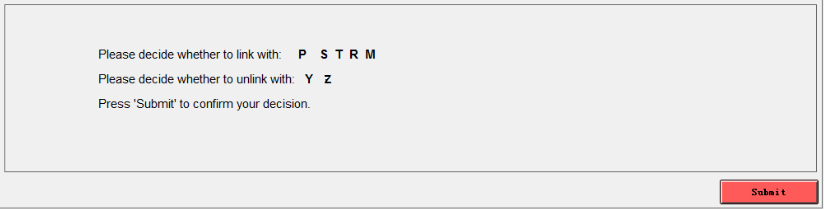}
\end{figure}

\begin{center}
	\bf Stage 2 - Yes/No Decision
\end{center}
In case you decided not to link to another participant in Stage 1 but this other participant has opted to send you a link proposal, then in Stage 2 we ask you to confirm your earlier decision not to link to this other participant. You will be shown the following table where you need to respond to link proposals in Stage 2.

Please refer to the attachment (Stage 2) for illustration purpose.

\begin{figure}[H]
	\centering
	\includegraphics[scale=0.8]{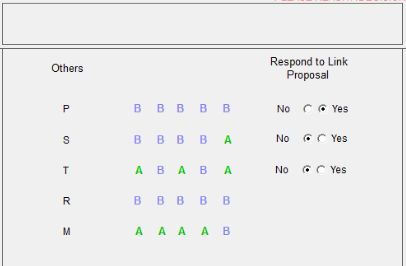}
\end{figure}

As in Stage 1, the table has 3 columns, from left to right:
\begin{itemize}
	\setlength\itemsep{0.0001em}
	\item {\bf ``Others''}: lists the IDs of the participants who are not linked with you
	\item {\bf ``Past actions''}: lists the last 5 actual actions by each of these participants
	\item {\bf ``Respond to link proposal''}: allows you to accept or reject a link proposal from a participant that is not currently linked with you
\end{itemize}

For example, look at participants P, S and T: under the ``Respond to link proposals'' column a ``No'' and a ``Yes'' buttons appear. This means that P, S and T have sent you a link proposal. For each proposal, you need to click on ``Yes'' to accept it or ``No'' to reject it. In the example shown, you have chosen to reject the proposals from S and T, and accept the proposal from P.

Note that if you have sent a proposal to connect to a participant in Stage 1 and that participant has done the same then the link is automatically formed without the need of further approval in Stage 2, and hence no button will be shown.		

\begin{center}
	\bf Stage 3 - Action Choice
\end{center}
In Stage 3, you choose either action \textcolor[rgb]{0.00,0.50,0.25}{A} or action \textcolor[rgb]{0.00,0.00,1.00}{B}. Please refer to the attachment (Stage 3) for illustration purpose.

\underline{Bottom-right of the screen}

This is where you choose your action in Stage 3 by clicking either the \textcolor[rgb]{0.00,0.50,0.25}{A} or \textcolor[rgb]{0.00,0.00,1.00}{B} button. You must pick an action in order to proceed.

\begin{figure}[H]
	\centering
	\includegraphics[scale=0.7]{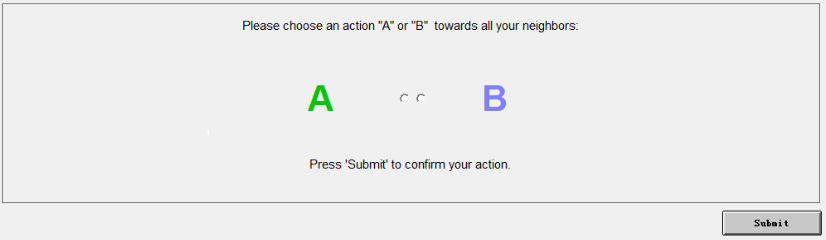}
\end{figure}

Note that it is possible that you are not linked with any participant (i.e. you have no neighbor) as the result of yours and others' earlier decisions. In such case, you will instead see the message reminding you that you do not have any neighbor and so you do not need to choose an action.

\begin{center}
	\bf Action Outcome
\end{center}
After Stage 3, you will see a summary of the points you got with each of the other participants in the round.

The action you have chosen ({\bf intended action}) during Stage 3 may or may not be implemented correctly. Specifically, there is a {\bf 15\% chance} that your intended action \textcolor[rgb]{0.00,0.50,0.25}{A} is implemented as {\bf actual action} \textcolor[rgb]{0.00,0.00,1.00}{B} and a {\bf 15\% chance} that your intended action \textcolor[rgb]{0.00,0.00,1.00}{B} is implemented as {\bf actual action} \textcolor[rgb]{0.00,0.50,0.25}{A}. Note that other players only observe your implemented ({\bf actual}) action and {\bf not} your {\bf intended} action.

Please refer to the attachment (Action Outcome) for illustration purpose.

The table shows your neighbors and your neighbors' {\bf actual actions} in this round. The first row of this table reminds you of your intended action and the actual action implemented. For example, in the table below your intended action was \textcolor[rgb]{0.00,0.50,0.25}{A} and your actual action was \textcolor[rgb]{0.00,0.50,0.25}{A}. Note that the table is updated with yours and the others' decisions in the previous Stages.

\begin{figure}[H]
	\centering
	\includegraphics[scale=0.7]{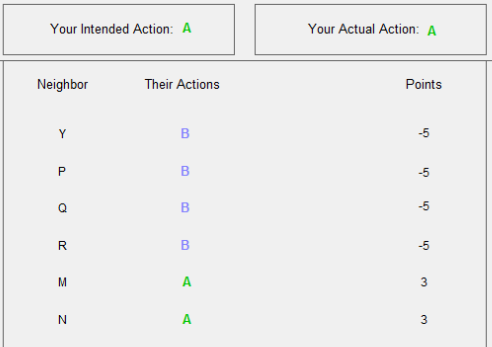}
\end{figure}

The table has 3 columns, from left to right:
\begin{itemize}
	\setlength\itemsep{0.0001em}
	\item {\bf ``Neighbor''}:         lists the IDs of the participants who are currently linked with you
	\item {\bf ``Action''}:           lists the actual action by each of your neighbors in this round
	\item {\bf ``Points''}:           lists the points you gained from your and each neighbor's actual action
\end{itemize}		

If your intended action is implemented correctly, you will see the following box.

\begin{figure}[H]
	\centering
	\includegraphics[scale=0.7]{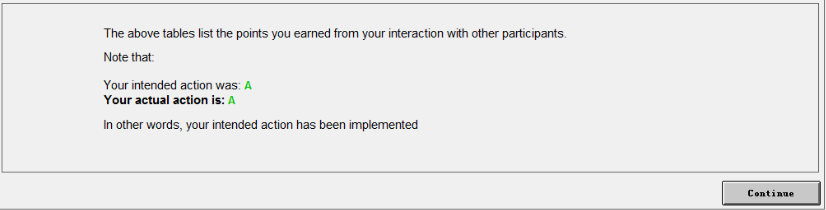}
\end{figure}

If your intended action is {\bf not} implemented correctly, you will see the following box. You need to click the ``Continue'' button to move to Stage 1 of the next round.

\begin{figure}[H]
	\centering
	\includegraphics[scale=0.7]{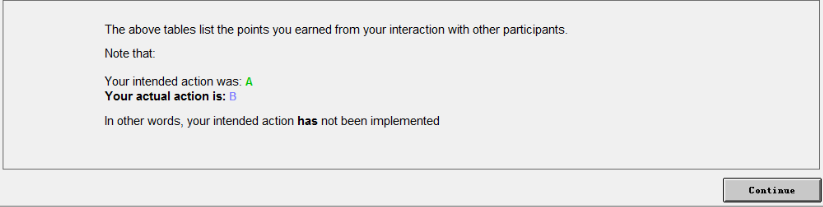}
\end{figure}

\begin{center}
	\underline{Earnings in Part I}
\end{center}
At the end of the Experiment, we will randomly select {\bf 6} rounds for payment. In each of these {\bf 6} rounds, we will randomly pick {\bf 2} of your potential link pairs (There are 11 potential link pairs in total). Note that a participant who was not linked with you can be picked, and in that case you will get 0 points for the interaction with that participant.

To determine your earnings, we sum the number of points you got with each of the picked participants in each of the {\bf 6} rounds. The exchange rate from ECU to SGD is:
\begin{center}
	\em 1.5 ECU = 1 SGD
\end{center}		
\noindent\rule{\textwidth}{0.4pt}
\begin{center}
	\bf Part II
\end{center}

In this part of the study you will be asked to make a series of choices. How much you receive will depend partly on chance and partly on the choices you make. The decision problems are not designed to test you. What we want to know is what choices you would make in them.

For each line in the table you will see in this stage, please indicate whether you prefer option A or option B. There will be a total of 10 lines in the table but just one line will be randomly selected for payment. You do not know which line will be paid when you make your choices. Hence you should pay attention to the choice you make in every line.

After you have completed all your choices, the computer will randomly generate a number, which determines which line is going to be paid out.

Your {\bf earnings for the selected line depend on which option you chose}: If you chose {\bf option A} in that line, you will receive {\bf 20 ECU}. If you chose {\bf option B} in that line, you will receive {\bf either 60 ECU or 0}. To determine your earnings in the case you chose option B, there will be a second random draw. The computer will randomly determine if your payoff is 0 or 60, with the chances set by the computer as they are stated in Option B.

\begin{center}
	\underline{Earning in Part II}
\end{center}
In Part II, the conversion rate from experimental dollars (ECU) to Singaporean dollars is:
\begin{center}
	\em 20 ECU = 1 SGD
\end{center}

\clearpage

\begin{figure}[H]
	\centering
	\includegraphics[scale=0.35]{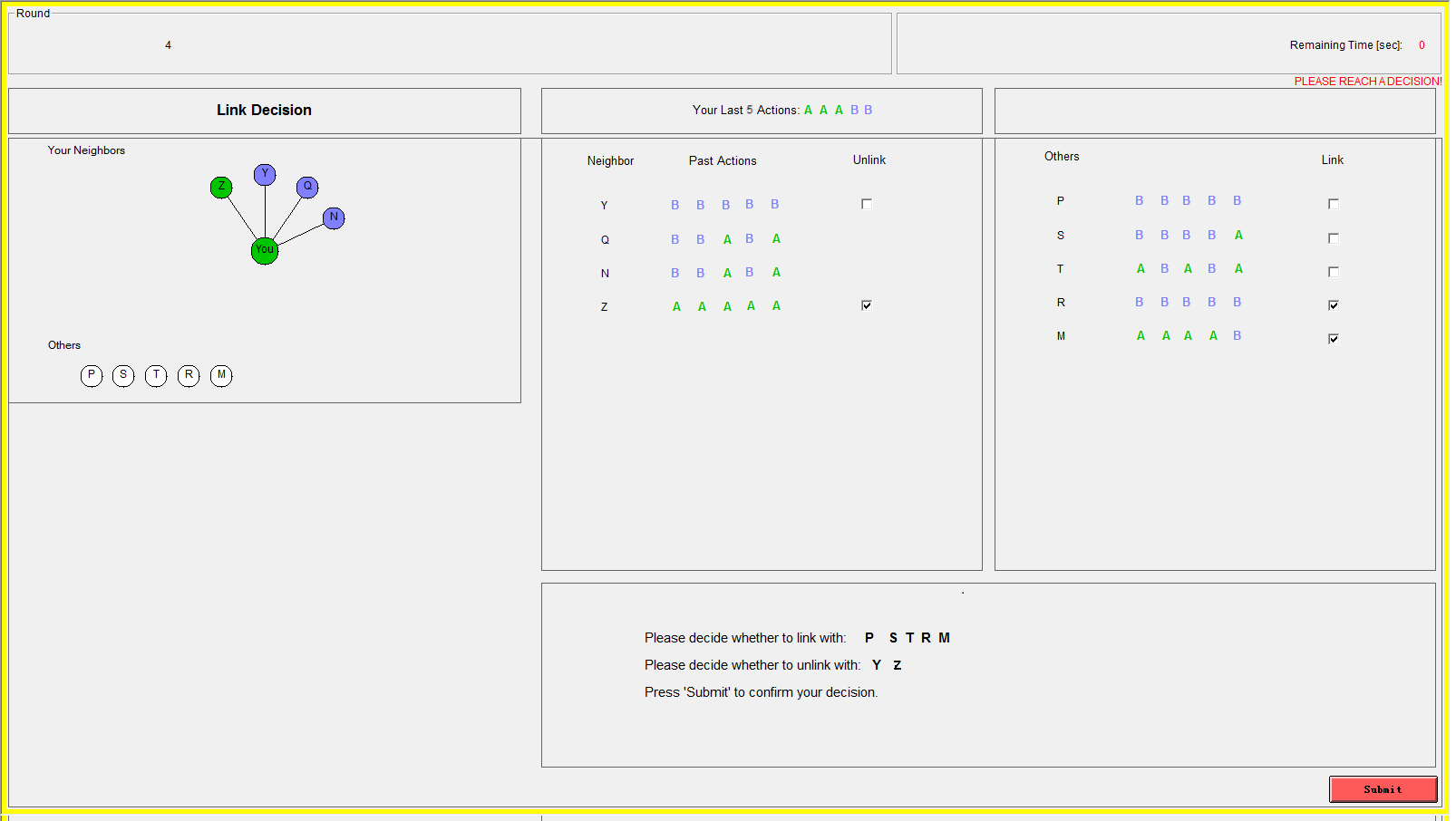}
	\caption*{Attachment (Stage 1)}
\end{figure}

\begin{figure}[H]
	\centering
	\includegraphics[scale=0.35]{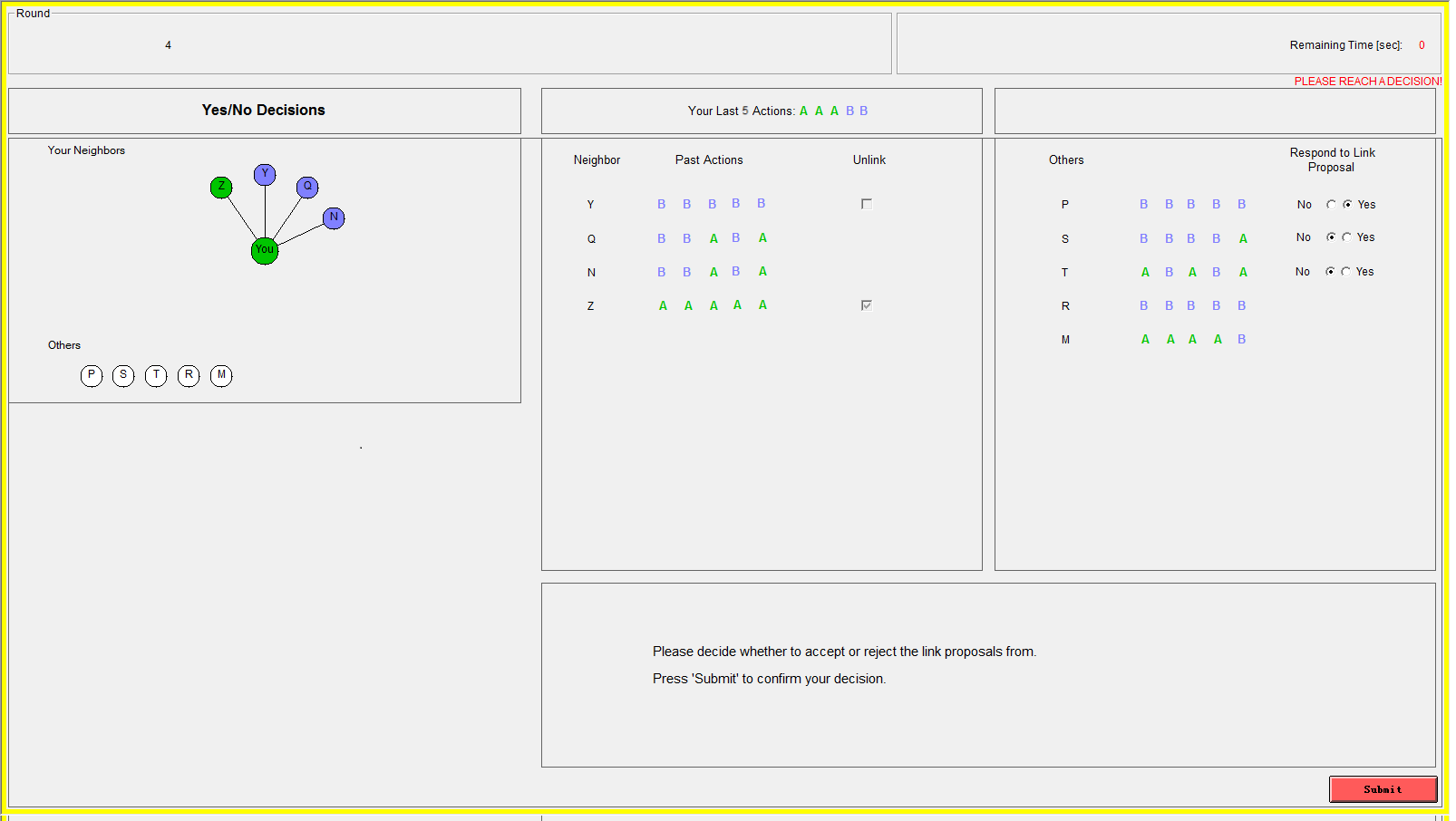}
	\caption*{Attachment (Stage 2)}
\end{figure}

\clearpage
\begin{figure}[H]
	\centering
	\includegraphics[scale=0.35]{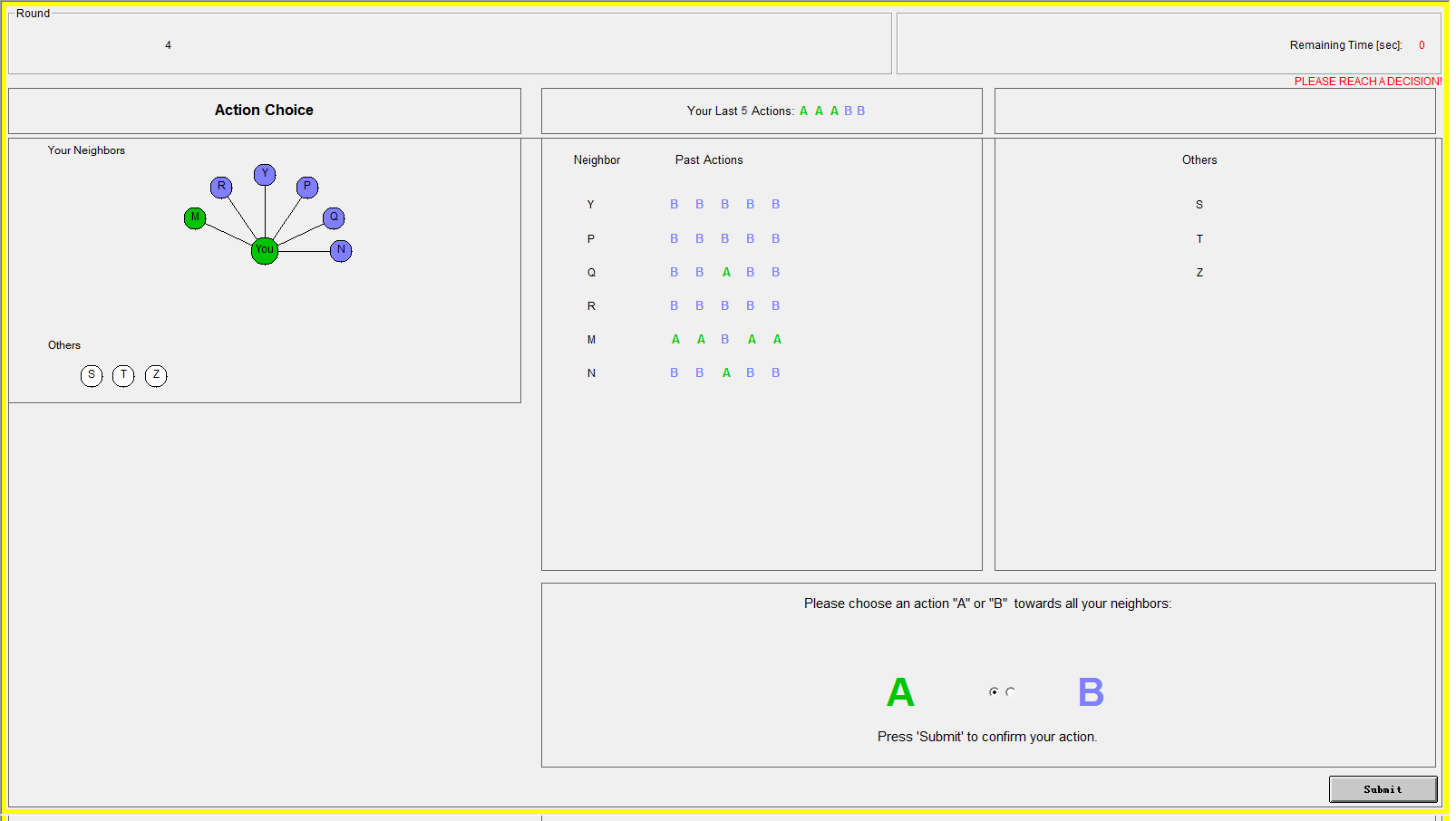}
	\caption*{Attachment (Stage 3)}
\end{figure}

\begin{figure}[H]
	\centering
	\includegraphics[scale=0.35]{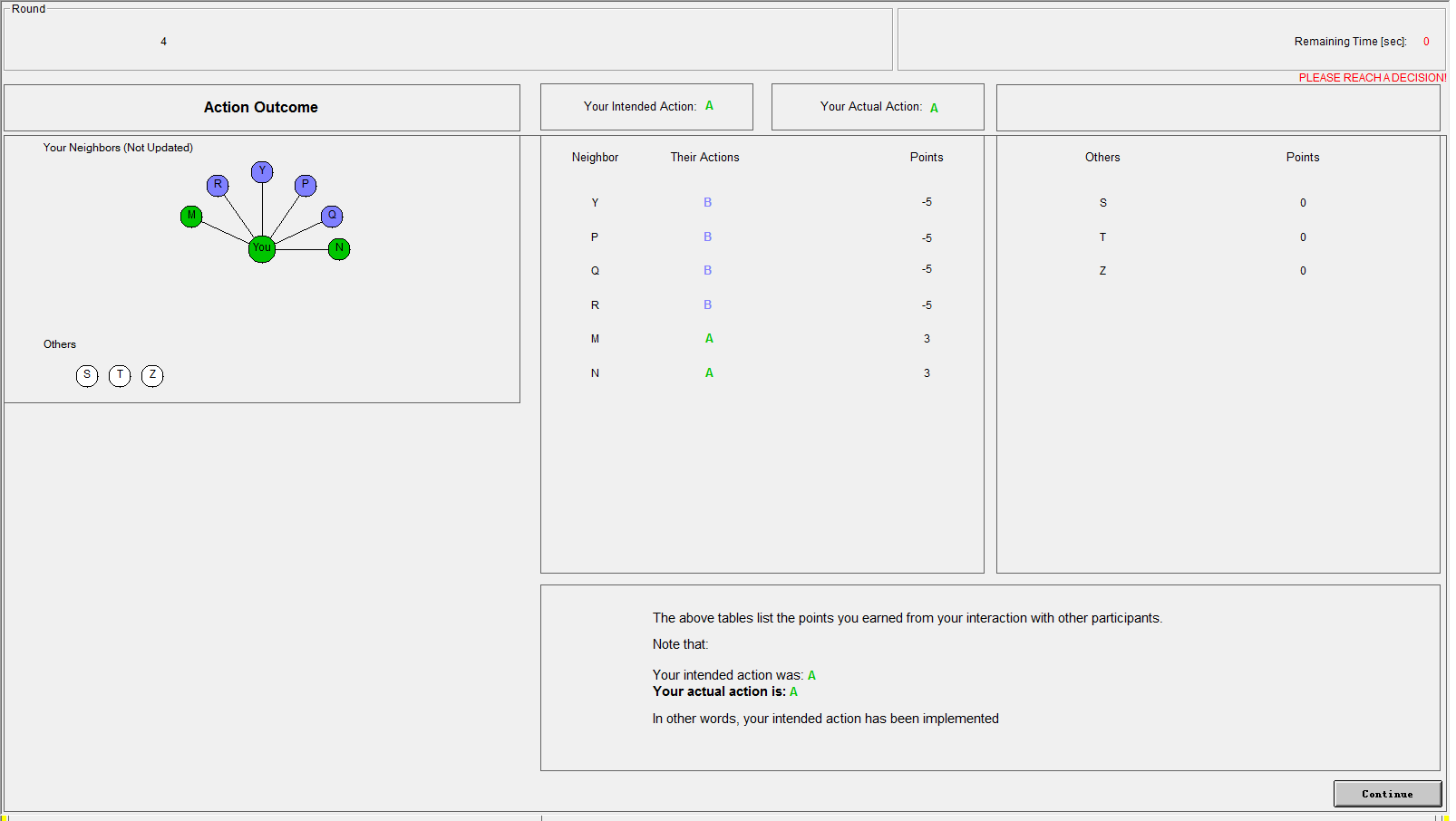}
	\caption*{Attachment (Action Outcome)}
\end{figure}

\end{small}

\end{spacing}

\end{document}